\documentclass[
preprint,
5p,
twocolumn,
12pt,
a4paper,
sort&compress,
]{elsarticle}
\usepackage{lipsum}
\usepackage{braket}
\usepackage{cancel}
\usepackage{tikz, pgf}
\usetikzlibrary{positioning}
\usepackage{graphicx}
\usepackage{amssymb, amsmath}
\usepackage{color, xcolor}
\definecolor{darkgreen}{rgb}{0.1,0.4,0.0}
\usepackage{hyperref}
\usepackage{times}
\usepackage{mathrsfs}
\usepackage[utf8]{inputenc}
\usepackage[capitalise]{cleveref}
\usepackage[normalem]{ulem}
\crefrangelabelformat{equation}{(#3#1#4--#5#2#6)}
\graphicspath{{./figs/}}
\newcommand{\rcite}[1]{\cite{#1}}
\newcommand{\rref}[1]{\rcite{#1}}
\newcommand{\refref}[1]{Ref.~\rcite{#1}}

\newcommand{\eref}[1]{Eq.~(\ref{#1})}
\newcommand{\sref}[1]{Sec.~\ref{#1}}

\newcommand{\fref}[1]{Fig.~\ref{#1}}

\newcommand{\dkt}{d^3 k\,}

\renewcommand{\bar}[1]{\mkern 1.8mu\overline{\mkern-1.8mu#1\mkern-1.8mu}\mkern 1.8mu}
\usepackage{bm} 
\newcommand{\vect}[1]{\bm{#1}}
\DeclareMathOperator{\atan}{atan}
\journal{arXiv}
\begin{document}
\begin{frontmatter}
\title{Effects of multiple single-particle basis states in scattering systems}
\author[CSSM]{Curtis D. Abell\corref{cor1}}
\author[CSSM]{Derek B. Leinweber}
\author[CSSM]{Anthony W. Thomas}
\author[UCAS,SCNT]{Jia-Jun Wu}
%
\cortext[cor1]{Corresponding author: {curtis.abell@adelaide.edu.au}}
\affiliation[CSSM]{
  organization={Special Research Centre for the Subatomic Structure of Matter (CSSM)},
  addressline={Department of Physics, University of Adelaide},
  city={Adelaide},
  postcode={5005},
  state={South Australia},
  country={Australia}}
\affiliation[UCAS]{
  organization={School of Physical Sciences},
  addressline={University of Chinese Academy of Sciences (UCAS)},
  city={Beijing},
  postcode={100049},
  country={China}}
\affiliation[SCNT]{
  organization={Southern Center for Nuclear-Science Theory (SCNT)},
  addressline={Institute of Modern Physics, Chinese Academy of Sciences},
  city={Huizhou},
  postcode={516000},
  state={Guangdong Province},
  country={China}}
%
%
%
\begin{abstract}
  Low-lying baryon resonances have been explored using Hamiltonian Effective Field Theory (HEFT), in a formalism where resonances with a three-quark component are described by both two-particle meson-baryon states and a bare basis state.
  Here, we investigate the use of multiple bare states in the Hamiltonian, to extend the formalism to higher energy ranges, and represent a larger portion of the low-lying baryon spectrum.
  Introducing a second bare state into a toy model extension of the low-energy $\Delta(1232)$ system, we explore the influence of the second bare state on the position of poles in the infinite-volume $T$-matrix.
  Considering the same system in a finite-volume, we analyse the finite-volume energy spectrum in the presence of a second bare state, providing insight into the interplay between two bare basis states, representing quark-model states, and the relationship between infinite-volume poles and finite-volume eigenstates.
\end{abstract}
\end{frontmatter}
%
%
%
%
\section{Introduction}
\label{sec:intro}
Given recent progress in the field, lattice QCD is now the prime method for studying low-lying baryon resonances in the context of non-perturbative QCD. However, as lattice QCD is formulated in terms of a Euclidean metric in a finite volume, scattering observables such as phase shifts and inelasticities are not directly obtainable.
L\"uscher's formalism \rref{Luscher:1985dn,Luscher:1986pf,Luscher:1990ux} is the widely-used method of relating the finite-volume quantities from lattice QCD with infinite-volume observables.
Though this process is exact for a simple scattering system with only a single scattering channel, generalisations to multiple channels~\rref{He:2005ey,Lage:2009zv,Bernard:2010fp,Guo:2012hv,Hu:2016shf,Li:2012bi,Hansen:2012bj}, including three-particle scattering~\rref{Doring:2018xxx,Hansen:2019nir,Blanton:2019vdk}, make the procedure rather complicated.
An alternative approach to the problem is Hamiltonian Effective Field Theory (HEFT), a Hamiltonian extension of effective field theory which provides a methodology for relating finite and infinite volume quantities. This relationship is guaranteed by L\"uscher's formalism, which is embedded within
HEFT~\rref{Wu:2014vma}.

In HEFT, we build a Hamiltonian consisting of two-particle basis states, and single particle bare basis states, where the interactions between these basis states are parametrised by separable potentials. The generalization to three-body channels is available via the relativistic Faddeev equations \rref{Thomas:1977ph}.
The idea behind one-particle bare basis states is that they contribute to baryons with a strong three-quark, single-particle component, or alternatively a baryon which is quark-model like~\cite{Thomas:1982kv}.

In seeking to find a physical interpretation of the results of lattice QCD, we associate a bare state with a lattice QCD eigenstate which has been excited by a three-quark interpolating field.
A particularly important feature of HEFT is the ability to extract the finite-volume eigenvectors of the Hamiltonian, which then provide insight into the structure of energy eigenstates on the lattice and the system as a whole.

Previous studies of baryon resonances utilising HEFT have been limited to Hamiltonians containing only a single bare basis state, and considering only a single resonance in the infinite-volume observables.
This has seen considerable success in studies of the $\Delta(1232)$ \rref{Hall:2013qba,Abell:2021awi}, the Roper $N^{*}(1440)$ \rref{Liu:2016uzk,Wu:2017qve}, the $\Lambda^{*}(1405)$ \rref{Hall:2014uca}, and the $N^{*}(1535)$ \rref{Liu:2015ktc}.
On the other hand, the spectrum of baryon resonances is considerably more complicated than a single resonance in each partial wave channel and future studies will require a multiple bare state formalism to fully explore the breadth of hadron structure.

In order to investigate the effects of a second bare basis state in both infinite-volume and finite-volume systems, we will consider a simple (''toy model'') extension of the $\Delta(1232)$ system considered in~\refref{Abell:2021awi}.
By limiting the investigation to the extension of a well-studied system, we aim to gain intuition into the behaviour of the position of poles associated with these bare states  in infinite volume, as well as the potential changes in the finite-volume energy spectrum.

In \sref{sec:theory}, we provide an overview of the HEFT formalism, with a particular emphasis on the nuances associated with multiple bare states.
This formalism will then be considered in the context of an extension of a study of the $\Delta(1232)$, where \sref{sec:inf} considers the infinite-volume behaviour of poles in the $T$-matrix as the second bare state is introduced.
A similar study is then performed in \sref{sec:fin}, where we consider the impact of a second bare state on the finite-volume energy spectrum.
In particular, the relationship between infinite and finite-volume behaviour is explored.
We conclude this report with a summary of the results in \sref{sec:conclusion}.
%
%

\section{Theoretical Framework}
\label{sec:theory}
\subsection{Hamiltonian Method}
\label{subsec:Hamiltonian}
Following the formalism produced in \refref{Abell:2021awi}, we commence with a description of the extension of the HEFT framework to include multiple bare states.
In the rest frame, the Hamiltonian for the system can be written as the sum of a non-interacting, free Hamiltonian, $H_0$, and an interaction Hamiltonian $H_{\text{I}}$,
\begin{equation}
  H = H_0 + H_{\text{I}}\,.
\end{equation}
The non-interacting Hamiltonian consists of two types of states.
First we have the free meson-baryon two-particle states with back-to-back momentum $\vect{k}$, labelled $\ket{\alpha(\vect{k})}$, where $m_{\alpha_\text{M}}$ and $m_{\alpha_\text{B}}$ are the masses of the mesons and baryons, respectively.
The energy associated with such a state is therefore given by $\omega_{\alpha}(k) = \sqrt{m_{\alpha_\text{M}}^2 + k^2} + \sqrt{m_{\alpha_\text{B}}^2 + k^2}$.
The second type are the bare single-particle basis states, which one may think of as representing quark model-like states.
These states are labelled $\ket{B_0}$, with unphysical basis-state mass given by $m_{B_0}$.
The free Hamiltonian for such a system can be expressed as
\begin{align}
  H_0 &= \sum_{B_0} \ket{B_0} m_{B_0} \bra{B_0} \nonumber\\
  &\quad+ \sum_{\alpha} \int\dkt\, \ket{\alpha(\vect k)} \omega_{\alpha}(\vect{k}) \bra{\alpha(\vect k)}\,.
  \label{eq:H0}
\end{align}
The interaction Hamiltonian, $H_{\text{I}}$, is governed by interactions between the single-particle bare states and the two-particle basis states, described by a potential $G_{\alpha}^{B_0}(k)$, and between the two-particle basis states $\ket{\alpha(k)}$ and $\ket{\beta(k')}$, described by a potential $V_{\alpha\beta}(k,k')$.
This therefore takes the form
\begin{align}
  H_{\text{I}} &= \sum_{B_0}\sum_{\alpha}\int d^3k\,\Bigl\{ \ket{B_0}G_{\alpha}^{B_0}(\vect k)\bra{\alpha(\vect k)} \Bigr. \nonumber\\
  &\quad + \Bigl. \ket{\alpha(\vect k)}{G_{\alpha}^{B_0}}^\dagger(\vect k)\bra{B_0} \Bigr\} \nonumber \\
  &\quad + \sum_{\alpha,\,\beta}\int d^3k\,\int d^3k'\, \ket{\alpha(\vect k)}V_{\alpha\beta}(\vect{k},\vect{k'})\bra{\beta(\vect k')}\,. \label{eq:HI}
\end{align}
In particular, we require a potential separable in the incoming and outgoing momenta.
While this formalism generalises potentials which are the sum of separable terms,
as in \refref{Hall:2014uca}, for the purposes of this study we will utilise a potential of the form $V_{\alpha\beta}(k,k') = v_{\alpha\beta} f_{\alpha}(k) f_{\beta}(k')$.

To remove the ultraviolet divergences in the Hamiltonian, we employ finite-range regularisation (FRR) \rref{Thomas:2002sj,Young:2002ib,Young:2002cj}.
FRR regulates the interaction strength at momentum $k$ through the introduction of a form-factor, $u(k)$, with a regulator parameter $\Lambda$.
It is desirable to have a regulator form which removes contributions from large momentum in a smooth fashion.
A dipole or Gaussian form factor is typically chosen to satisfy this condition, and as such a dipole regulator of the form
\begin{equation}
  u(k) = \left( 1 + \frac{k^2}{\Lambda^2} \right)^{-2}\,,
\end{equation}
is used for this study.
Furthermore, for a simple single-channel system, this independence is extended to the choice of regulator parameter for values ranging from $\Lambda = 0.6$ to $\Lambda = 8.0$ GeV.
For more complex systems, with multiple two-particle channels, experimental data constrains the regulator parameter to $\Lambda \sim 1$ GeV \rref{Abell:2021awi}.
%
%

\subsection{Infinite-Volume Scattering} \label{subsec:inf}
By calculating the $T$ matrix for this Hamiltonian, we are able to solve for the $P$-wave phase shifts and constrain the free parameters in the Hamiltonian.
We are also able to search for the position of any poles in the $T$ matrix to investigate the properties of any resonances or bound states that may appear in the system.
For a system with multiple bare basis states, the reduced coupled-channel scattering equations take the form
\begin{align}
  T_{\alpha\beta}&(k,k';E) = \tilde{V}_{\alpha\beta}(k,k';E) \nonumber\\
  &+ \sum_{\gamma}\, \int dq\, q^2\, \frac{\tilde{V}_{\alpha\gamma}(k,q;E)\, T_{\gamma\beta}(q,k';E)}{E - \omega_{\gamma}(q) + i\epsilon}\,,
\end{align}
where the sum over $\gamma$ considers all channels, each with centre-of-mass
energy $\omega_{\gamma}(q)$.
Here, we have defined the coupled-channel potential $\tilde{V}_{\alpha\beta}$, which takes the form
\begin{equation}
  \tilde{V}_{\alpha\beta}(k,k';E) = V_{\alpha\beta}(k,k') + \sum_{B_0} \frac{G_{\alpha}^{B_0\, \dagger}(k)\, G_{\beta}^{B_0}(k')}{E - m_{B_0}}\,,
\end{equation}
where this potential sums over all bare states $B_0$.

As this study is focussed primarily on the interplay of bare states, we turn our attention to the portion of the $T$-matrix which describes contributions from the bare states.
In principle, poles also arise from the portion of the $T$-matrix which describes interactions between two-particle basis states.
As shown in \refref{Abell:2021awi} however, without a bare basis state we require an unphysical regulator parameter to produce poles in the vicinity of the physical $\Delta(1232)$ pole position.
Thus, the relevant portion of the $T$-matrix takes the form
\begin{equation} \label{eq:bare_Tmatrix}
  T_{\alpha\beta}^{\text{bare}}(k,k';E) = \sum_{B_0,B'_0}\, \mathcal{G}_{\alpha}^{B_0\, \dagger}(k)\, A_{B_0,B'_0}(E)\, \mathcal{G}_{\beta}^{B'_0}(k') \, ,
\end{equation}
where $A_{B_0,B'_0}(E)$ is the dressed propagator
\begin{equation}
  A_{B_0,B'_0}(E) = \left[ \delta_{B_0,B'_0}\, (E - m_{B_0}) - \bar{\Sigma}_{B_0,B'_0}(E) \right]^{-1} \, . \label{eq:Amat}
\end{equation}
Here, $\bar{\Sigma}_{B_0,B'_0}(E)$ represents the sum of all one-loop self-energy interactions.
An example of this is the transition $\ket{B_0} \rightarrow \ket{B'_0}$ via intermediate scattering channels $\ket{\gamma(k)}$, given by
\begin{equation} \label{eq:self_energy}
  \Sigma_{B_0,B'_0}(E) = \sum_{\gamma} \int dq\, q^2\, \frac{G_{\gamma}^{B_0\, \dagger}(q)\, G_{\gamma}^{B'_0}(q)}{E - \omega_{\gamma}(q) + i\epsilon} \, .
\end{equation}

In \eref{eq:bare_Tmatrix}, $\mathcal{G}_{\alpha}^{B_0}(k)$ is a modified potential which describes how the bare state is dressed by the background interactions between two-particle basis states.
Poles associated with the bare states are generated purely in the propagator $A_{B_0,B'_0}(E)$, and can therefore be found by searching for complex energies such that $\det\left( A_{B_0,B'_0}(E)^{-1} \right) = 0$.
This equation emphasises how the two bare states are mixed by the two-particle interactions in forming the physical poles.
%
%

\subsection{Finite-Volume Hamiltonian}
In order to generate a finite-volume energy spectrum and make a connection to lattice QCD results, we constrain the Hamiltonian to a three-dimensional cube of volume $L^3$.
Within this volume, the momentum is discretised to $\vect{k}_{\vect{n}} = 2\pi \vect{n}/L$, where $\vect{n} \in \mathbb{Z}^{3}$.
This set of allowed momenta can be simplified by considering degenerate momentum states, where the degeneracy for some $n = |\vect{n}|^2$ is described by $C_{3}(n)$.
This is a function which counts the number of degenerate momentum states for each $n$, and therefore under this discretisation scheme, integrals over all $k$-space reduce from
\begin{equation}
  \int\, \dkt\, \rightarrow \left(\frac{2\pi}{L}\right)^{3}\, \sum_{\vect{n}\in\mathbb{Z}^3} = \left(\frac{2\pi}{L}\right)^{3}\, \sum_{n=1}^{n_{\text{max}}}\, C_{3}(n) \, ,
\end{equation}
where $n_{\text{max}}$ is chosen such that the regulator $u(k)$ is sufficiently small \rref{Abell:2021awi}.
As momentum is discretised, and reaches a maximimum for $k_{n_{\max}} = 2\pi n_{\text{max}}/L$, the Hamiltonian can be written as a finite matrix, for which the eigenvalues and eigenvectors can be solved.
In particular, as the eigenvectors describe the linear combination of basis states for a given eigenvalue, they provide insight into the composition of corresponding states calculated in lattice QCD.
In this discretisation scheme, the potentials $G_{\alpha}^{B_0}(k)$ and $V_{\alpha\beta}(k,k')$, as defined in \sref{subsec:Hamiltonian}, take on finite-volume factors of the form
\begin{align}
  \bar G_{\alpha}^{B_0}(k_n) &= \sqrt{\frac{C_3(n)}{4\pi}}\, \left( \frac{2\pi}{L} \right)^{\frac{3}{2}}\, G_{\alpha}^{B_0}(k_n)\,, \nonumber\\
  \bar V_{\alpha\beta}(k_n, k_{m}) &= \sqrt{\frac{C_3(n)}{4\pi}}\, \sqrt{\frac{C_3(m)}{4\pi}}\, \left( \frac{2\pi}{L} \right)^{3} \\
                            &\qquad\qquad \times V_{\alpha\beta}(k_n, k_m) \,.
\end{align}

While the form of the Hamiltonian matrix for a single bare baryon state has been covered extensively in previous work \rref{Hall:2013qba,Hall:2014uca,Liu:2015ktc,Liu:2016uzk,Wu:2017qve,Li:2019qvh,Liu:2020foc,Li:2021mob,Abell:2021awi,Guo:2022hud}, the multiple bare baryon state scenario is yet to be explored.
The addition of a second bare state requires only an additional row and column for the Hamiltonian.
Then the free Hamiltonian $H_0$ has the form
\begin{align}
  H_0 = \text{diag}&\left( m_{B_0}\,, m_{B'_0}\,, \omega_{\alpha}(k_0)\,, \right. \nonumber\\
        &\quad\qquad \left. \omega_{\alpha}(k_1)\,, \omega_{\alpha}(k_2)\,, \dots\,, \omega_{\alpha}(k_{n_{\max}}) \right)\,.
\end{align}
For a system with two bare states and a single scattering channel, as will be considered in this report, we may exclude the index referring to the scattering channel.
In this case, the interaction Hamiltonian, $H_{\text{I}}$, can be written as
\begin{align}
  &H_{\text{I}} = \nonumber\\
    &\begin{pmatrix}
    0 & 0 & \bar G^{B_0}(k_1) & \bar G^{B_0}(k_2) & \cdots \\
    0 & 0 & \bar G^{B'_0}(k_1) & \bar G^{B'_0}(k_2) & \cdots \\
    \bar G^{B_0}(k_1) & \bar G^{B'_0}(k_1)  & \bar V(k_1,k_1) &  \bar V(k_1,k_2) & \cdots \\
    \bar G^{B_0}(k_2) & \bar G^{B'_0}(k_2)  & \bar V(k_2,k_1) &  \bar V(k_2,k_2) & \cdots \\
	\vdots & \vdots & \vdots & \vdots & \ddots
  \end{pmatrix}
\end{align}
As the Hamiltonian matrix is finite, standard eigenvalue equation methods can be used to solve for the finite-volume spectrum and eigenvectors.
%
%

\section{Two bare states in infinite volume} \label{sec:inf}
In order to concentrate this study on the effects of introducing a second bare state to the Hamiltonian, we begin with the system defined in \refref{Abell:2021awi}, where using a single bare state, and only a $\pi N$ scattering channel, the phase shifts are extracted from the $T$-matrix.
These phase shifts are fit to scattering data provided by SAID~\rref{site:SAID,Workman:2012hx}, up to the $\pi\Delta$ threshold of approximately 1.35 GeV.
In this formalism, the interaction between the bare $\Delta$ and a $\pi N$ two-particle basis state with momentum $k$ is given by
\begin{equation}
  G_{\pi N}^{\Delta}(k) = \frac{g_{\pi N}^{\Delta}}{m_{\pi}}\, \frac{k}{\sqrt{\omega_{\pi}(k)}}\, u(k)\,, \label{eq:Gk}
\end{equation}
where $u(k)$ is the dipole form factor, and $m_\pi$ is the physical pion mass.
Similarly, the interaction between $\pi N$ scattering states is given as
\begin{equation}
  V_{\pi N\pi N}(k,k') = \frac{v_{\pi N\pi N}}{m_{\pi}^2}\, \frac{k}{\omega_\pi(k)}\, \frac{k'}{\omega_\pi(k')}\,
u(k)\, u(k') \,.
\end{equation}
Using these potentials, and for a regulator parameter of $\Lambda = 0.8$ GeV, the fitting process results in a bare mass of $m_{\Delta_0} = 1.359$ GeV, and couplings $g_{\pi N}^{\Delta} = 0.176$, and $v_{\pi N\pi N} = -0.0127$.
Utilising the process outlined in \sref{subsec:inf}, a pole is found at the position
$E_{\text{pole}} = 1.211 - 0.049\,i$ GeV.
This is consistent with the Particle Data Group (PDG) value of $E_{\text{pole}} = 1.210 - 0.050\,i$ GeV \rref{ParticleDataGroup:2022pth}.

With this description of the experimental data, we are able to investigate how the introduction of a second bare state alters the system.
To do so, we distinguish the two bare states by labelling the first one used to fit data as $\Delta_1$, and the second as $\Delta_2$.
By smoothly turning on the coupling of the second bare state $g_{\pi N}^{\Delta_2}$, while holding all other parameters fixed, we are able to observe how the system is affected.
While this process does move away from the physical description of the $\Delta(1232)$ and we no longer have a good description of the scattering data, this instead allows us analyse exactly how the system is affected through the introduction of a second bare state.

In doing, so we consider two situations.
In the first, we introduce a second bare state with a larger bare mass.
This is likely to be the most common situation, where now we attempt to introduce bare basis states to represent excited states with a possible three-quark core.
In the second situation, we introduce a second bare basis state with a mass lower than the first bare state.
This may prove useful for studies of systems such as the Roper resonance, where a previous HEFT study found that a bare mass of 2.0 GeV was required to describe the finite-volume energy spectrum \rref{Wu:2017qve}.

\subsection{Large second bare mass} \label{subsec:inf_large}
\begin{figure}
  \centering
  \includegraphics[width=0.48\textwidth]{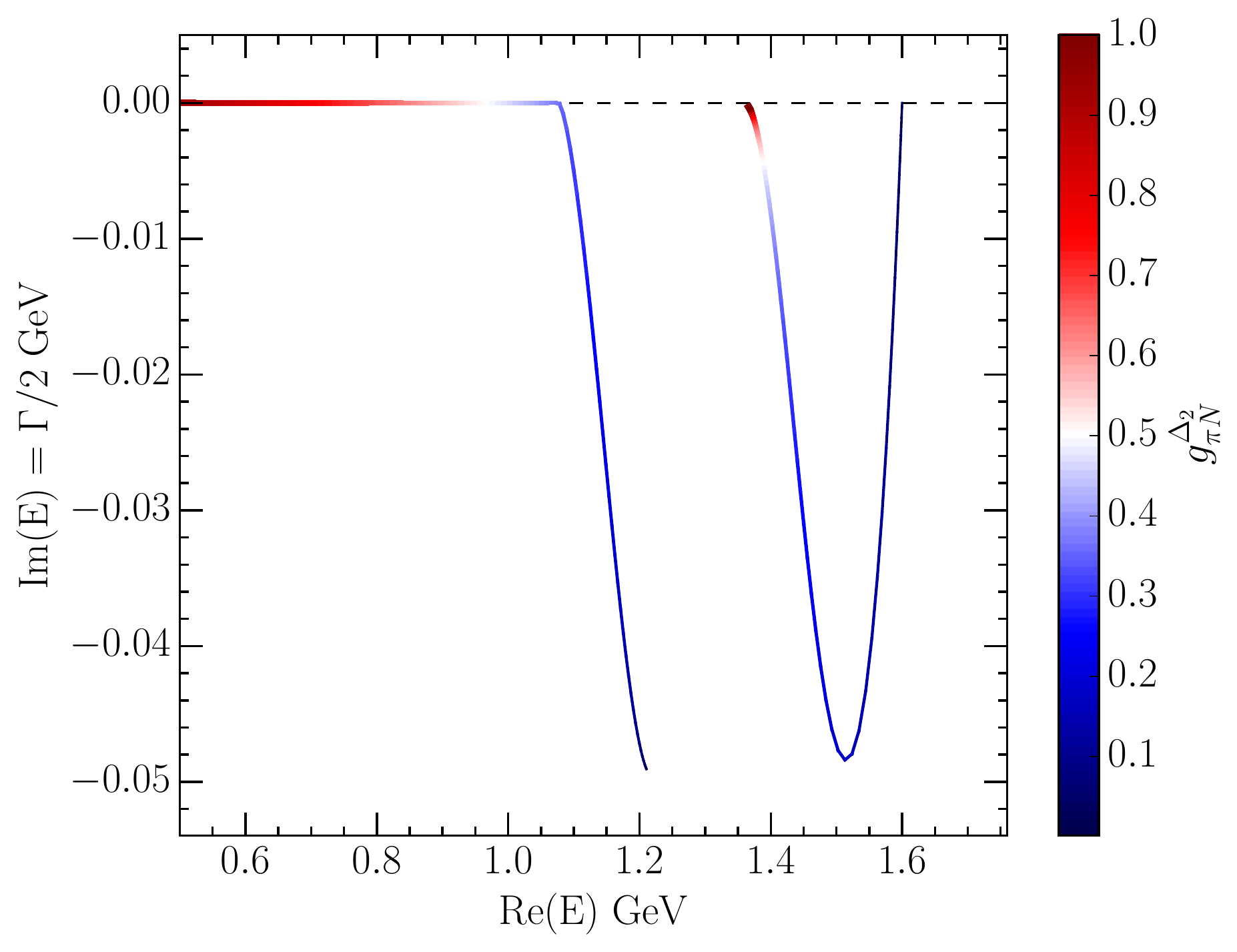}
  \caption{Variation of the $T$-matrix poles as the coupling of a second bare basis state to $\pi N$ scattering states is increased.
    The thin (blue) points represent the pole positions in the complex place for small $g_{\pi N}^{\Delta_2}$, moving to thicker (red) points for a larger coupling.
    Here, the mass of the second bare basis state, $\Delta_2$, is $1.6$ GeV, relative to $\Delta_1$ at 1.359 GeV.}
  \label{fig:2b1c_m16_poles}
\end{figure}

Considering this first system, we introduce a second bare state with a mass $m_{\Delta_2} = 1.6$ GeV.
Initially, the presence of this bare state with no coupling simply introduces a pole in the $T$-matrix on the real axis, at 1.6 GeV.
As the coupling $g_{\pi N}^{\Delta_2}$ is turned on, this non-interacting bound state gains a negative imaginary component, becoming a resonance as it decays to $\pi N$.
This process is demonstrated in \fref{fig:2b1c_m16_poles}, where we are smoothly increasing $g_{\pi N}^{\Delta_2}$ up to a value of 1.0, showing the movement of the new pole we have introduced, as well as the movement of the original pole representing the physical $\Delta(1232)$.

While initially located near the position of the physical $\Delta$ baryon, as $g_{\pi N}^{\Delta_2}$ increases the original pole rapidly tends to the real axis, while the second pole initially develops a larger imaginary component.
In the regime where $g_{\pi N}^{\Delta_1}$ and $g_{\pi N}^{\Delta_2}$ are of similar order, the two poles are both distinct and have reasonable widths.
Interestingly, the turning point in the position of the second pole occurs when $g_{\pi N}^{\Delta_1} = g_{\pi N}^{\Delta_2}$.
As the coupling of the second bare state begins moving to a value larger than that of the first, both poles tend towards the real axis.

Once the coupling of the second bare state has become approximately double that of the first, we find that the initial pole drops to a mass below the $\pi N$ threshold, becoming a bound state.
As this occurs, the pole corresponding to the second bare state also continues its trajectory towards the real axis.
Unlike the first pole however, as the second coupling increases, this pole only approaches the real axis, never becoming a bound state.
As the imaginary component of this pole approaches zero, the real component approaches the mass of the first bare state, at 1.359 GeV.
This intriguing result hints at an exchange in the composition of the physical states.
For small coupling, $\Delta_2$ is associated with the second pole, but at large coupling it is associated with the first pole, leaving the second pole to be governed by $\Delta_1$.

As described in detail in \ref{app:imag_poles}, the imaginary component of the larger mass pole is zero at $\mathcal{O}((g_{\pi N}^{\Delta_{1}}/g_{\pi N}^{\Delta_{2}})^{2})$, with the position of this pole given by $E = m_{\Delta_{1}} + (g_{\pi N}^{\Delta_{1}}/g_{\pi N}^{\Delta_{2}})^{2}(m_{\Delta_{2}} - m_{\Delta_{1}})$.
The imaginary components of this pole only arise at $\mathcal{O}((g_{\pi N}^{\Delta_{1}}/g_{\pi N}^{\Delta_{2}})^{4})$.

For studies involving multiple bare states, this pole movement implies that couplings between each bare state and scattering states, such as $g_{\pi N}^{\Delta_1}$ and $g_{\pi N}^{\Delta_2}$, should be of similar order to avoid the generation of unphysical poles and bound states.
It is only in this regime that the widths of each resonance correspond with typical particle widths observed in physical resonances.
In this final state, where $g_{\pi N}^{\Delta_2}$ is an order of magnitude larger than $g_{\pi N}^{\Delta_1}$, the system is completely dominated by $\ket{\Delta_2}$, and our system is unable to give a good description of resonance physics.
%
%
%

\subsection{Small second bare mass} \label{subsec_inf_small}
\begin{figure}
  \centering
  \includegraphics[width=0.48\textwidth]{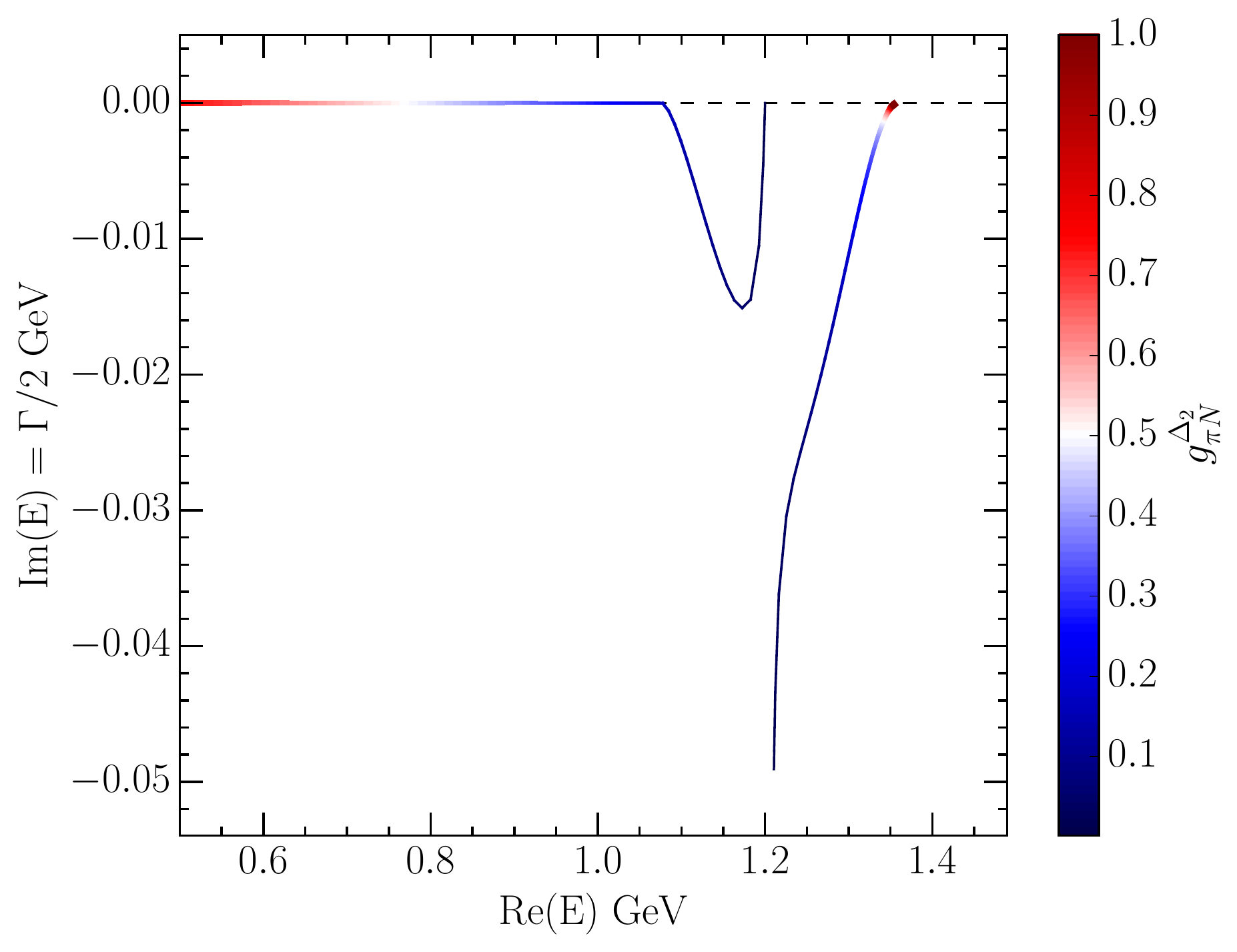}
  \caption{Variation of the $T$-matrix poles as the coupling of a second bare state to $\pi N$ scattering states is increased.
    The thin (blue) points represent the pole positions in the complex place for small $g_{\pi N}^{\Delta_2}$, moving to thicker (red) points for a larger coupling.
    Here, the mass of the second bare basis state, $\Delta_2$, is 1.2 GeV, relative to $\Delta_1$ at $1.359$ GeV.}
  \label{fig:2b1c_m1200_poles}
\end{figure}

As a second scenario we consider the addition of a second bare state with a mass lower than the original $m_{\Delta_1} = 1.359$ GeV.
In particular, we choose this mass to be $m_{\Delta_2} = 1.2$ GeV, as this lies approximately half way between the first bare mass and the $\pi N$ threshold, giving the clearest demonstration of novel behaviour.
While the introduction of a second bare state at $m_{\Delta_2} = 1.6$ GeV caused the original pole to move towards a bound state below the $\pi N$ threshold, introducing a second bare state with a mass less than the first has the opposite effect, as seen in \fref{fig:2b1c_m1200_poles}.
In this system, the initial behaviour is the same as the previous case, where we have a pole at the physical $\Delta(1232)$ position, and a second pole bound at $m_{\Delta_2} = 1.2$ GeV.
As $g_{\pi N}^{\Delta_2}$ increases in the same manner as \sref{subsec:inf_large}, where all other couplings are held fixed, the second pole initially behaves in the same manner, moving away from the real axis as it decays to $\pi N$.
The turning point occurs at $g_{\pi N}^{\Delta_1} = g_{\pi N}^{\Delta_2}$.
From here, however, it is the new second pole which continues to the $\pi N$ threshold, before becoming bound below this threshold.
The initial pole, which starts at the position of the physical $\Delta(1232)$, instead tends towards the real axis, at the position of the original bare state at $m_{\Delta_1} = 1.359$ GeV.

For large coupling of $\Delta_2$ to the $\pi N$ scattering states, the system consists of a bound state below the $\pi N$ threshold, and a long-lived state at the position of the original bare mass.
As described in \ref{app:imag_poles}, we find that the imaginary component of the larger mass pole goes like $(g_{\pi N}^{\Delta_{1}}/g_{\pi N}^{\Delta_{2}})^{2}$, with pole position $E = m_{\Delta_{1}} - (g_{\pi N}^{\Delta_{1}}/g_{\pi N}^{\Delta_{2}})^{2} (m_{\Delta_{1}} - m_{\Delta_{2}})$, consistent with the behaviour observed in \fref{fig:2b1c_m1200_poles}.
The imaginary components of this pole only arise at $\mathcal{O}((g_{\pi N}^{\Delta_{1}}/g_{\pi N}^{\Delta_{2}})^{4})$.

While the large-coupling behaviour of the two poles is the same regardless of the second bare state's mass, the path taken is reversed for the two situations considered.
This highlights the complexity of a system with multiple bare states, and the difficulty in identifying the origin of poles emerging from the interactions between multiple bare states.

While initially we have a pole which is identified as the physical $\Delta(1232)$ baryon, the introduction of a second bare state introduces a degree of ambiguity as to the origin and composition of the resultant poles, depending on the mass of the second bare state.
In order to gain a better understanding of the structure of these poles, as well as the behaviour of the system as the second bare state is introduced, we now turn our attention to finite-volume physics.
%
%

\section{Two bare states in a finite volume} \label{sec:fin}
Considering finite-volume physics allows one to gain additional insight into the behaviour and composition of the scattering states, through analysis of the Hamiltonian eigenvalues and eigenvectors,  respectively.
Using the same systems as in \sref{sec:inf}, we consider how the finite-volume spectrum is affected by the introduction of a second bare state.
Initially, with the second bare state turned off, the finite-volume energy spectrum is equivalent to the single-channel analysis in \refref{Abell:2021awi}, as well as with finite-volume energies calculated from L\"uscher's method.
For this study, we choose to concentrate on a lattice extent of $L = 3$ fm, giving a sufficient density of states to demonstrate the behaviour of the spectrum.

\subsection{Large second bare mass} \label{subsec:fin_large}
\begin{figure}
  \centering
  \includegraphics[width=0.48\textwidth]{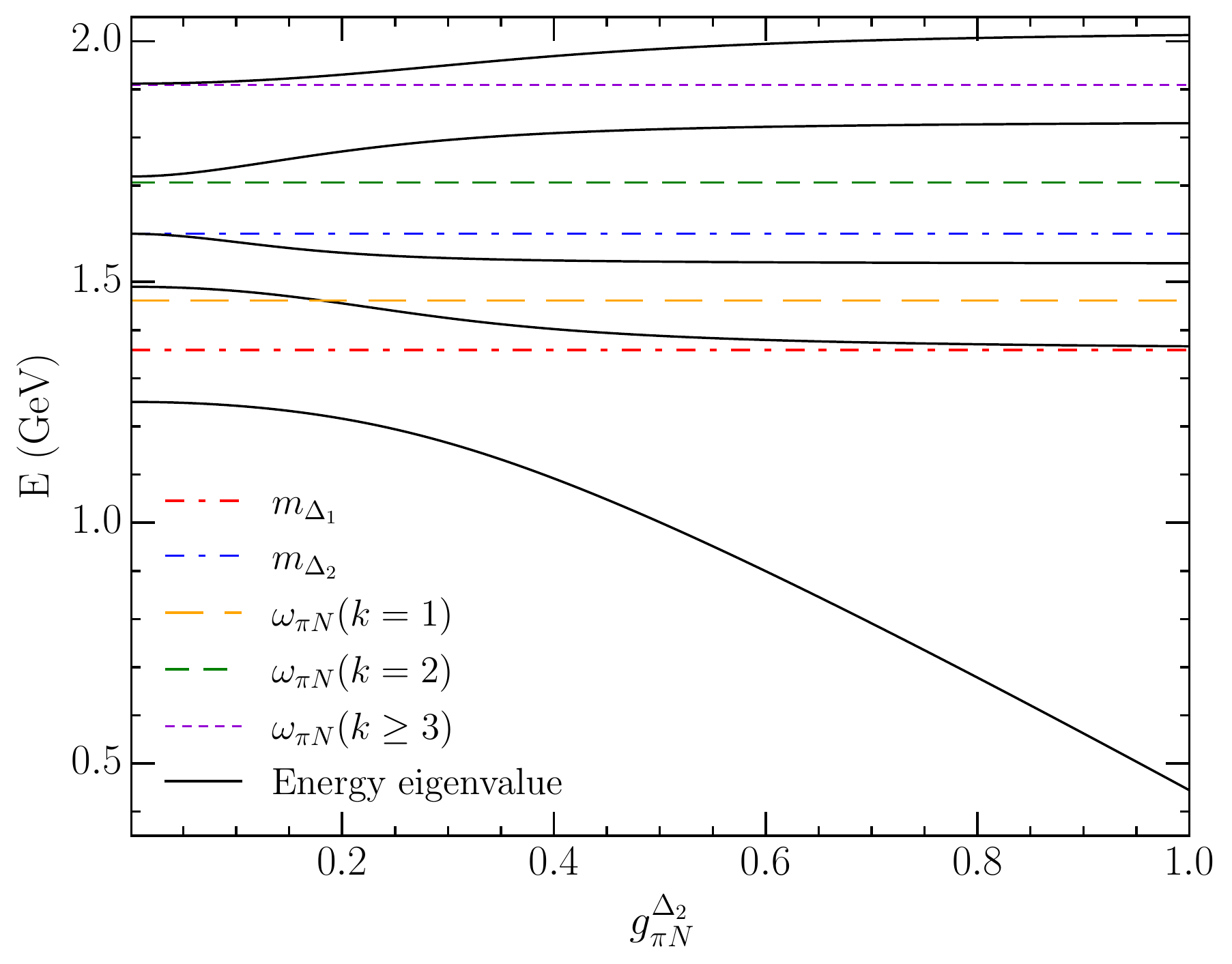}
  \caption{Dependence of the $L = 3$ fm finite-volume Hamiltonian energy eigenvalues on the coupling of the second bare basis state to $\pi N$ scattering states, labelled $g_{\pi N}^{\Delta_2}$.
    Here, the mass of the second bare basis state, $\Delta_2$, is 1.6 GeV, relative to $\Delta_1$ at $1.359$ GeV.}
  \label{fig:2b1c_EvG_L3fm}
\end{figure}
With the addition of a second bare basis state with mass $m_{\Delta_2} = 1.6$ GeV, the finite-volume spectrum shown in \fref{fig:2b1c_EvG_L3fm} displays similar behaviour to the poles
in \fref{fig:2b1c_m16_poles}.
Here, the third eigenvalue is initially identified at the position of the second bare mass, with no deflection from the non-interacting energy level.
As the coupling $g_{\pi N}^{\Delta_2}$ is increased, this eigenvalue shifts away from the non-interacting bare state at $1.6$ GeV, before tending towards the value of the lower-lying bare mass at $1.359$ GeV.
It is notable however that this movement begins at the third eigenvalue, before finishing at the second eigenvalue.
Also similar to the pole movement in \fref{fig:2b1c_m16_poles}, the lowest-lying energy eigenvalue shifts from a state near the position of the physical $\Delta(1232)$, to a bound state with energy less than the $\pi N$ threshold, at $m_{\pi} + m_{N} = 1.078$ GeV.
%
%
\begin{figure}
  \centering
  \includegraphics[width=0.36\textwidth]{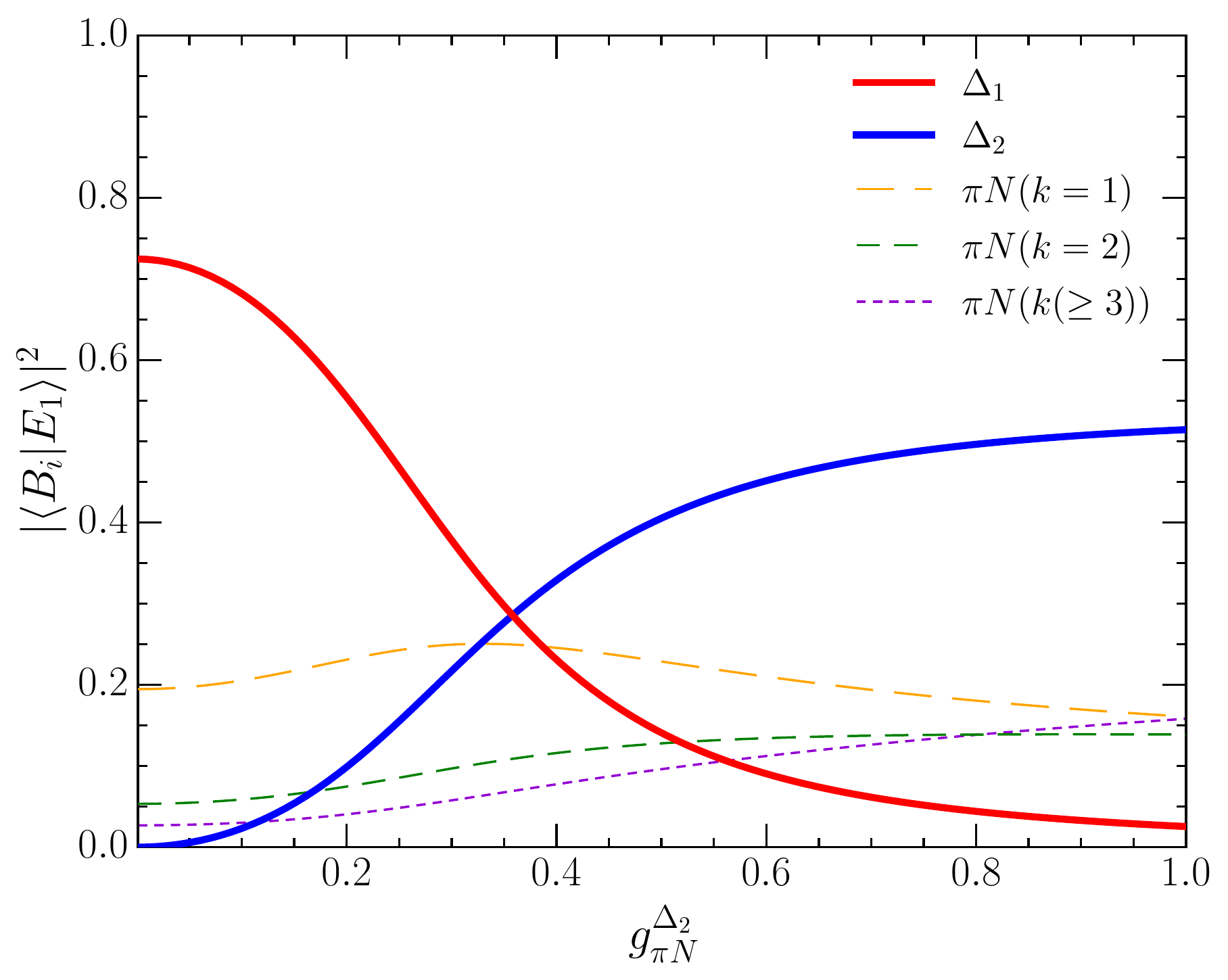}~\\
  \includegraphics[width=0.36\textwidth]{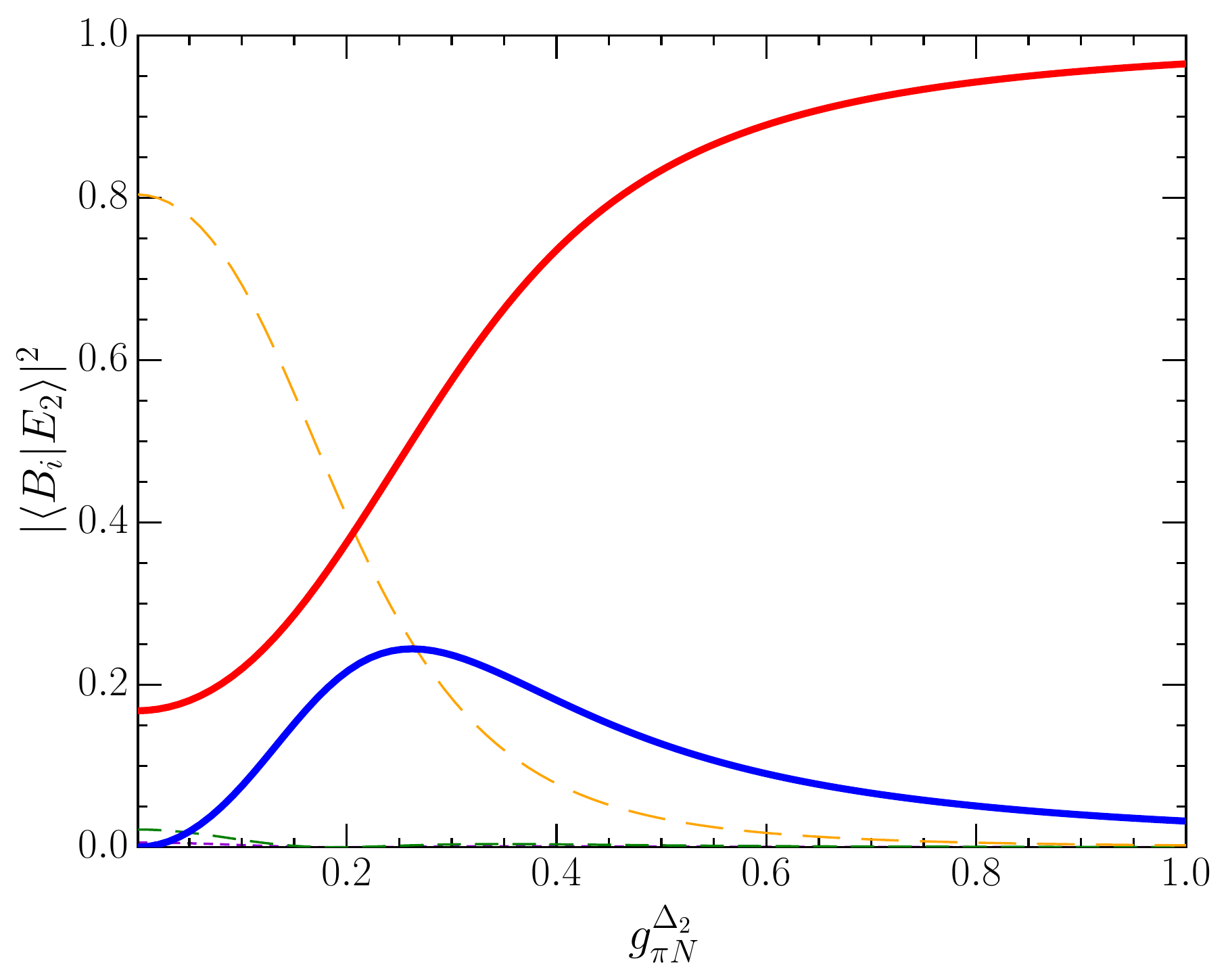}~\\
  \includegraphics[width=0.36\textwidth]{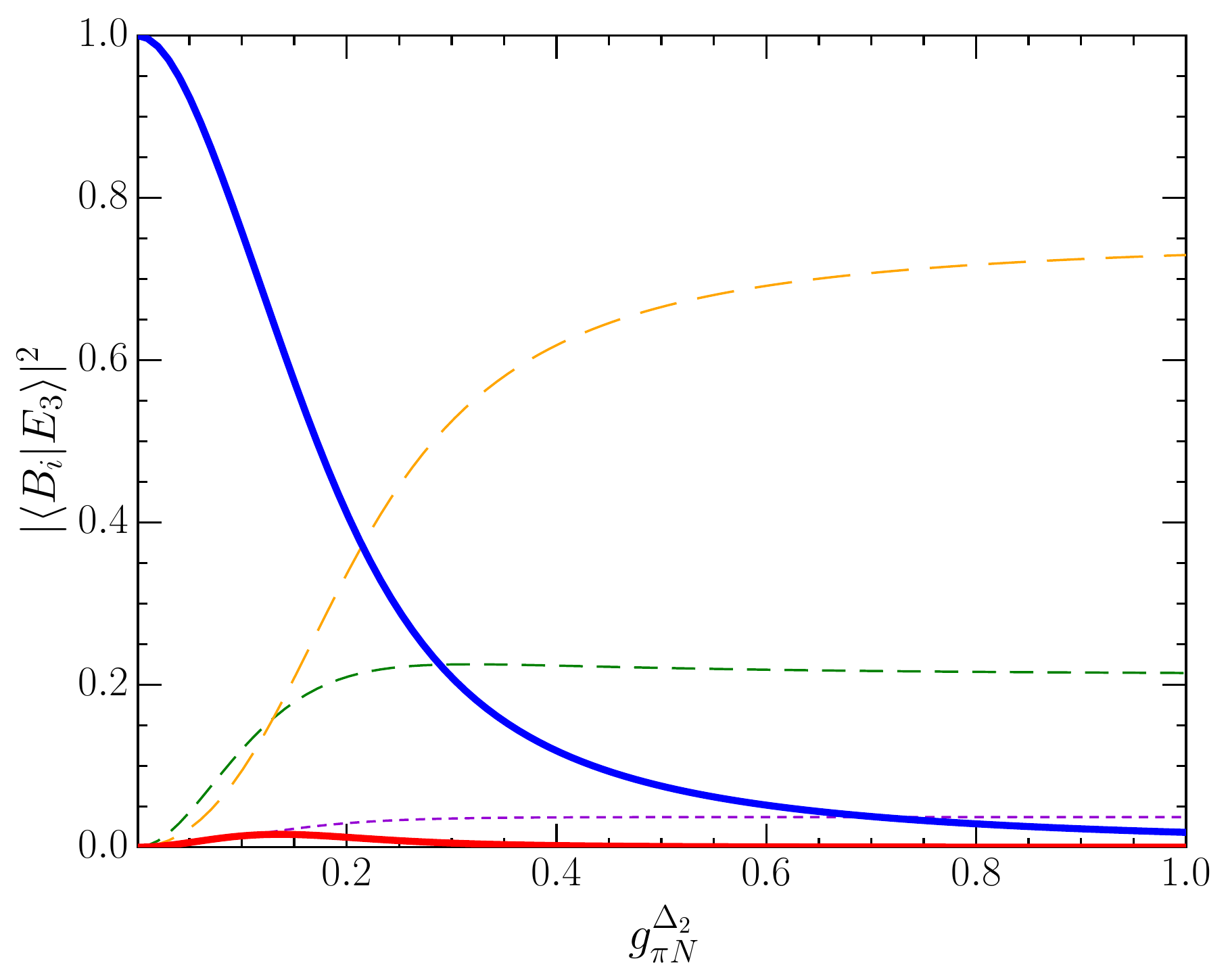}~\\
  \includegraphics[width=0.36\textwidth]{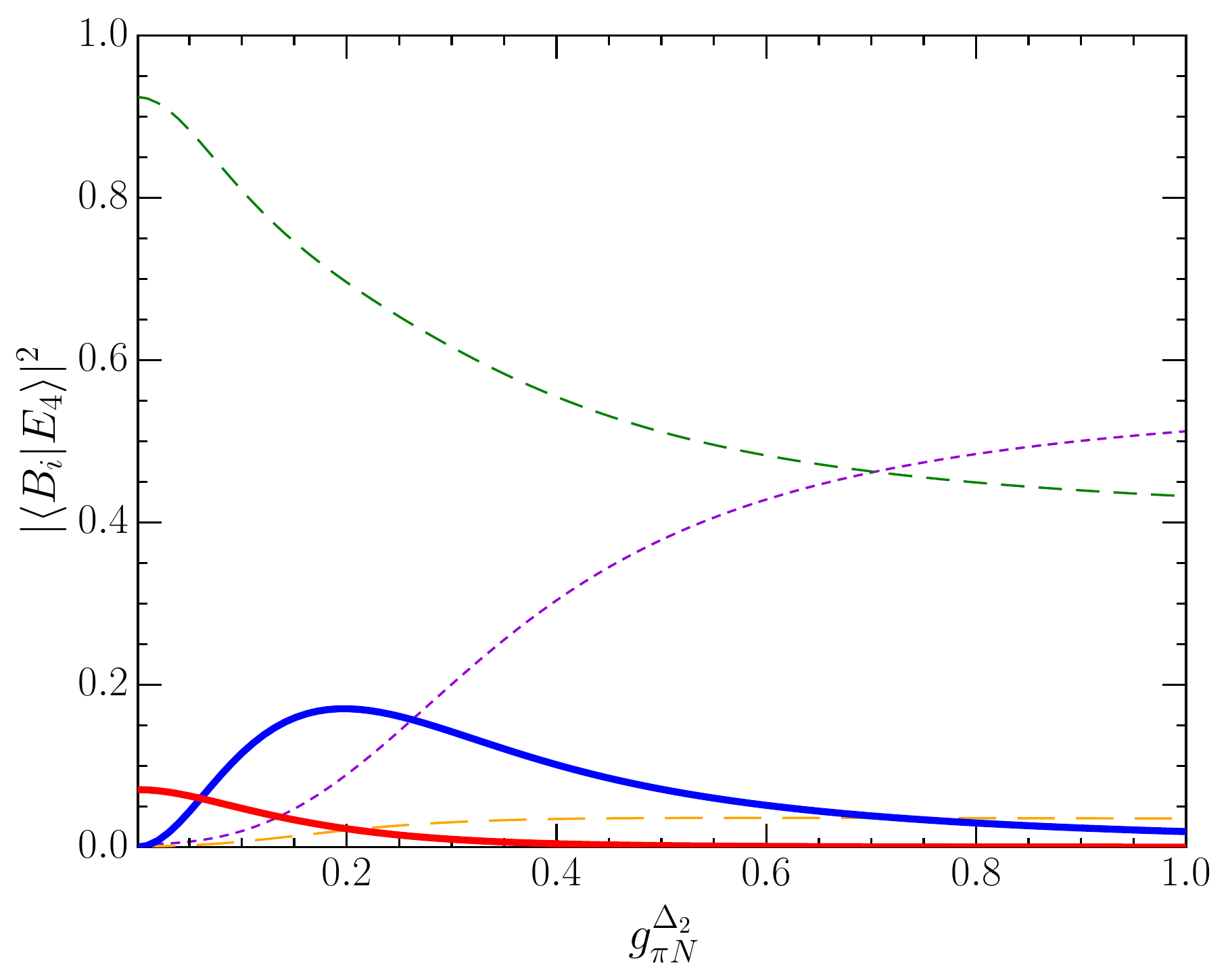}~
  \caption{Dependence of the finite-volume energy-eigenstate basis state structure on the coupling $g_{\pi N}^{\Delta_2}$, for $L = 3$ fm, where $m_{\Delta_2} = 1.6$ GeV.
    The label $\left| \bra{B_{i}}\ket{E_j} \right|^2$ denotes the contribution from each basis state $\ket{B_{i}}$ to the energy eigenvalues $\ket{E_{j}}$ from \fref{fig:2b1c_EvG_L3fm}.
  Eigenvector components from higher momentum scattering states are summed in $\pi N(k\geq 3)$.}
  \label{fig:2b1c_eigenvectors}
\end{figure}

An advantage of HEFT is the ability to extract the eigenvectors of the Hamiltonian, and gain insight into the contribution from basis states towards each eigenstate.
The dependence of the state composition on $g_{\pi N}^{\Delta_2}$ is shown in \fref{fig:2b1c_eigenvectors}.
Here we observe that, as the coupling to the second bare state increases, the interpretation of each eigenstate changes.
Initially, the ground state is dominated by contributions from the first bare state, while the second bare state contributions dominate the third eigenstate located at the mass of the second bare state.
As the coupling increases however, the state dominated by the second bare state quickly moves to lower energy, moving to the ground state for a coupling of $g_{\pi N}^{\Delta_2} \sim 0.4$.
It is only in the regime where $g_{\pi N}^{\Delta_1} = g_{\pi N}^{\Delta_2}$ where contributions from each bare state are spread through the tower of excited states.
At all coupling strengths we are able to distinctly identify a state in which the majority of contribution from a bare state is present.
Such a state is one which we expect to observe in the spectrum of lattice QCD generated by three-quark interpolating fields.

\begin{figure}
  \centering
  \includegraphics[width=0.48\textwidth]{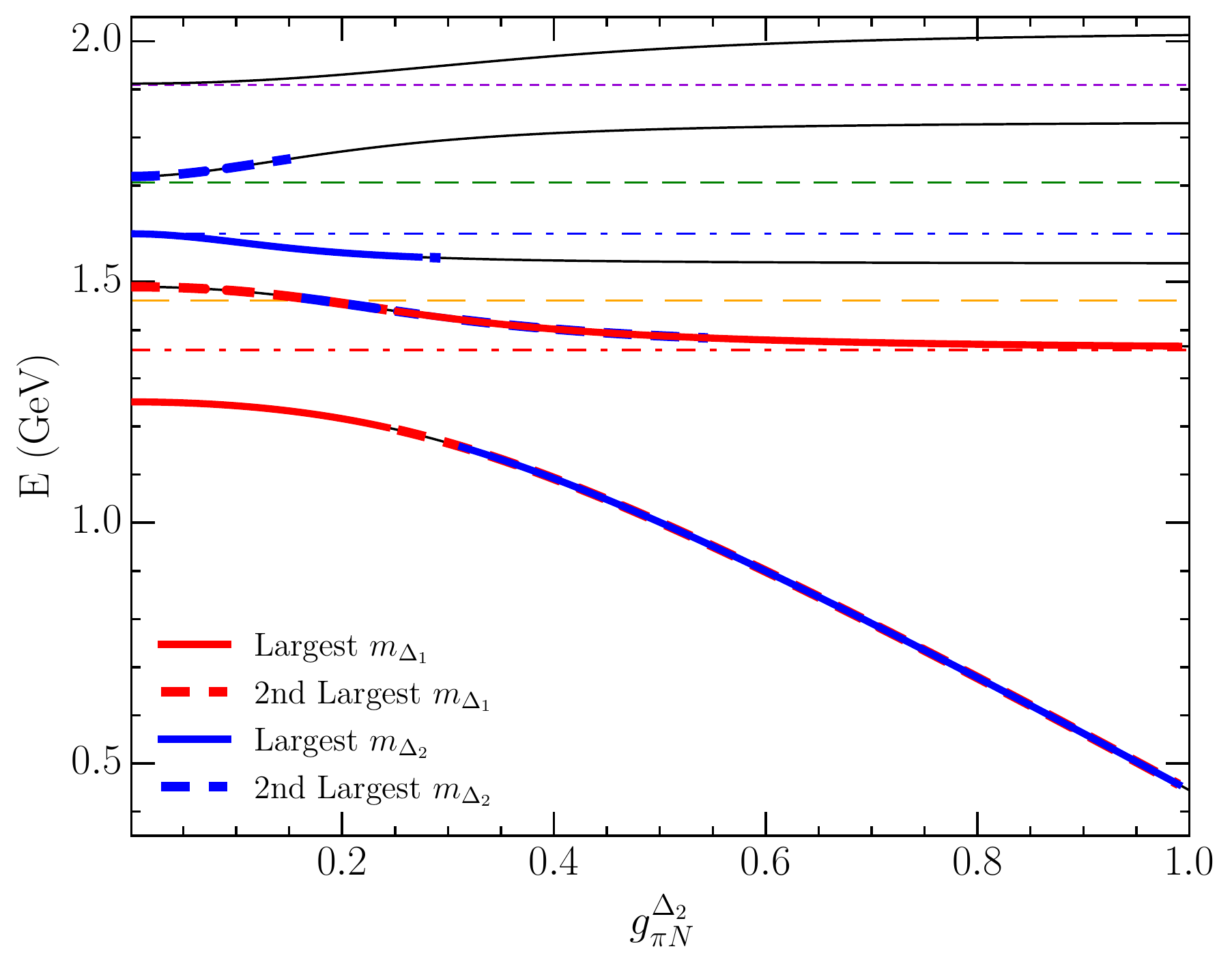}
  \caption{Dependence of the $L = 3$ fm finite-volume Hamiltonian energy eigenvalues on the coupling of the second bare basis state to $\pi N$ scattering states, labelled $g_{\pi N}^{\Delta_2}$.
    Here, the mass of the second bare basis state, $\Delta_2$, is 1.6 GeV, relative to $\Delta_1$ at $1.359$ GeV.
    Red and blue highlighting represents the states with largest eigenvector components from the first and second bare basis states respectively.
  Basis states are styled as in \fref{fig:2b1c_EvG_L3fm}.}
  \label{fig:2b1c_EvG_bare_L3fm_m1600}
\end{figure}

In order to better observe this distribution of contributions from each bare state, in \fref{fig:2b1c_EvG_bare_L3fm_m1600} we display the spectrum from \fref{fig:2b1c_EvG_L3fm} with additional information from the eigenvectors plotted, where the states with the first and second largest contributions from each bare state have been highlighted.
Using this scheme, it becomes clear how the position of the eigenstate dominated by each bare basis state moves as a function of $g_{\pi N}^{\Delta_2}$.
In particular, we are able to gain additional insight that is not obvious in the infinite-volume analysis.

While the behaviour of the eigenvalues matches the behaviour of the poles in \sref{subsec:inf_large}, the behaviour of the eigenvectors tells a new story.
At small coupling, $\Delta_2$ is associated with the third energy eigenstate.
However we see that it is the ground state which possesses the majority of $\ket{\Delta_2}$ at large coupling.
Thus $\Delta_2$ generates the lower lying pole at large coupling.
The exchange in the bare-state roles is made possible as $\Delta_1$ becomes associated with the second energy eigenstate.
At approximately $g_{\pi N}^{\Delta_1} = g_{\pi N}^{\Delta_2}$, the identities of the three energy eigenstates are shuffled.

\subsection{Small second bare mass} \label{subsec_fin_small}
\begin{figure}
  \centering
  \includegraphics[width=0.48\textwidth]{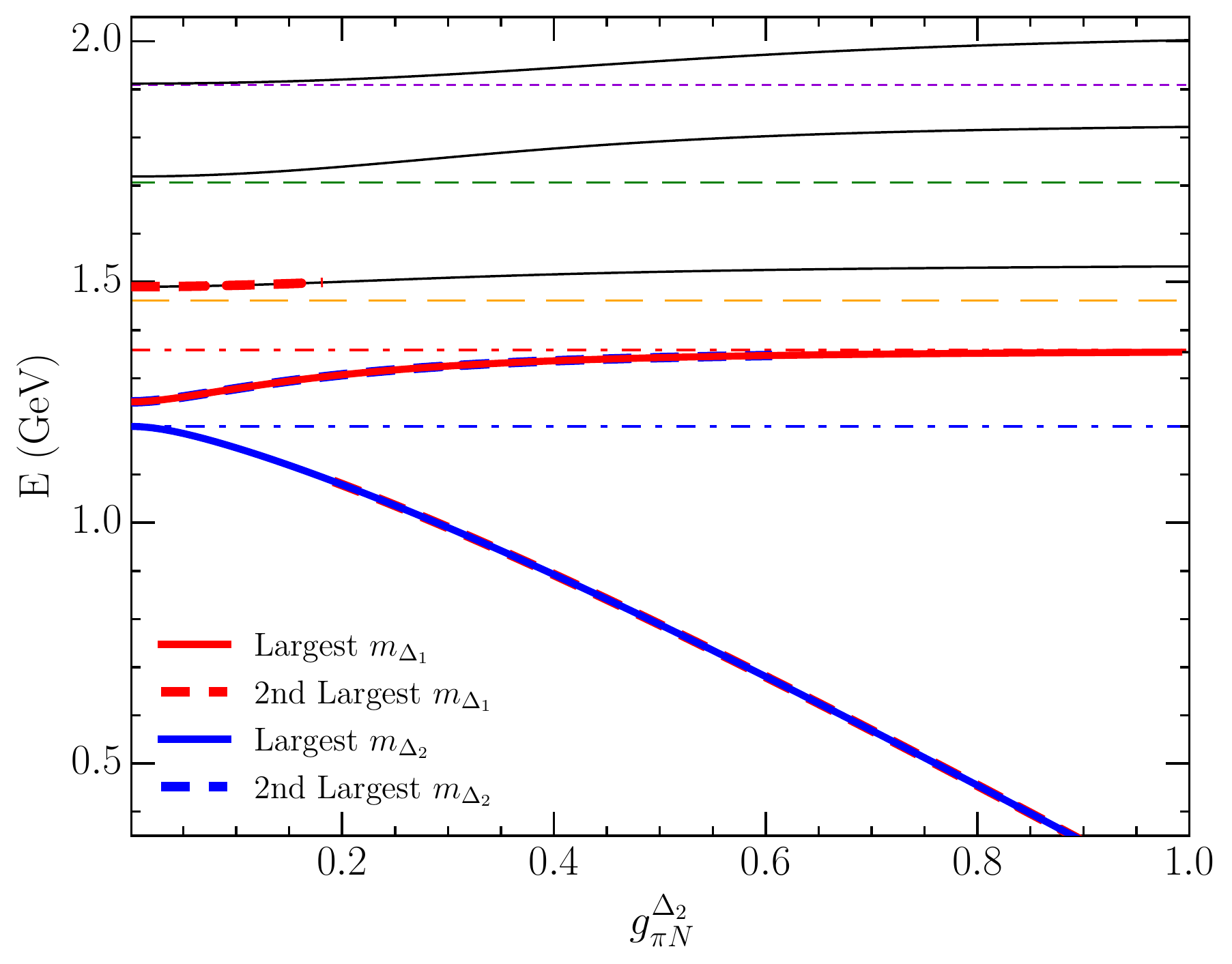}
  \caption{Dependence of the $L = 3$ fm finite-volume Hamiltonian energy eigenvalues on the coupling of the second bare basis state to $\pi N$ scattering states, labelled $g_{\pi N}^{\Delta_2}$.
    Here, the mass of the second bare basis state, $\Delta_2$, is 1.2 GeV, relative to $\Delta_1$ at $1.359$ GeV.
    Red and blue highlighting represents the states with largest eigenvector components from the first and second bare basis states respectively.
    Basis states are styled as in \fref{fig:2b1c_EvG_L3fm}.}
  \label{fig:2b1c_EvG_bare_L3fm_m1200}
\end{figure}

\begin{figure}
  \centering
  \includegraphics[width=0.36\textwidth]{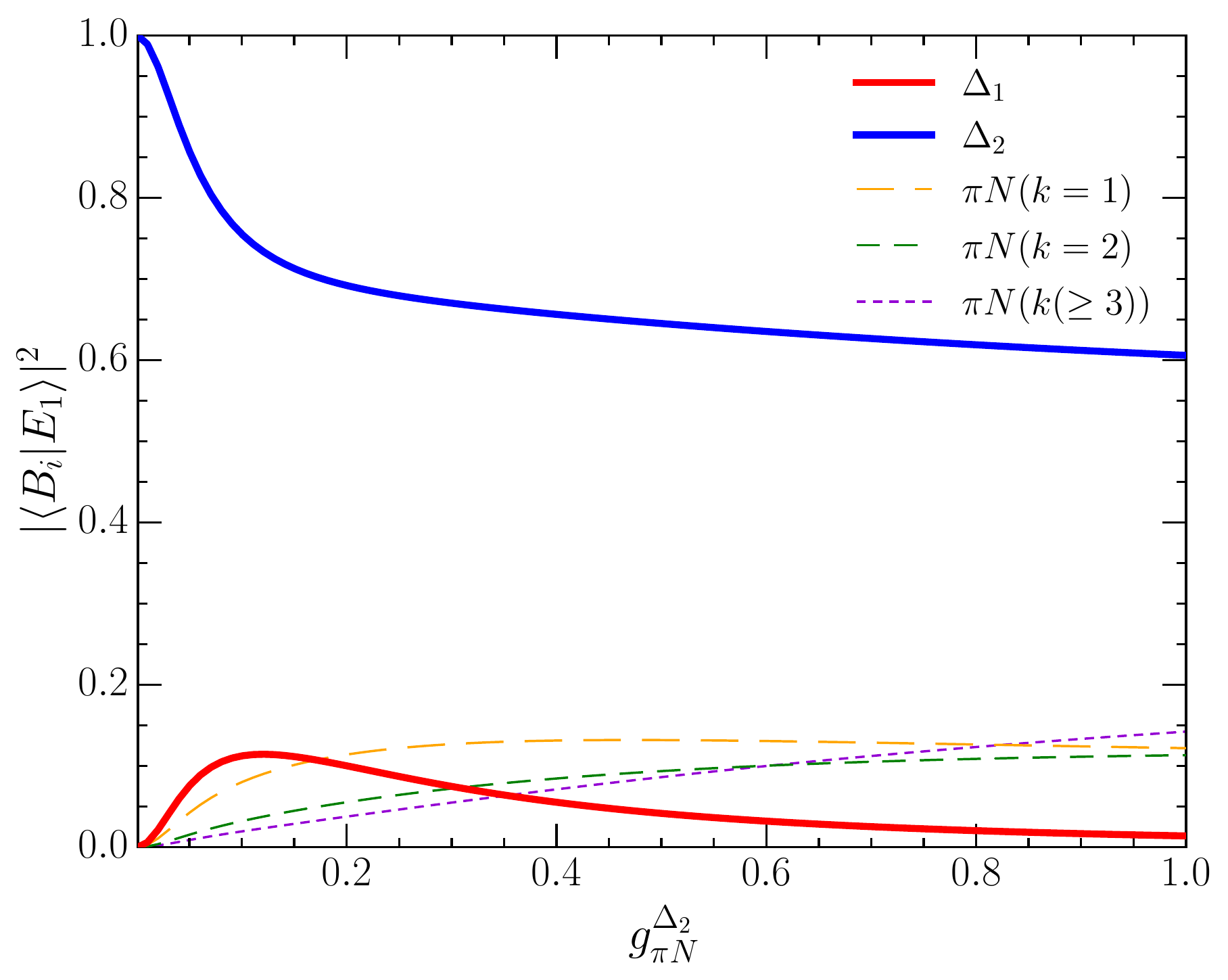}~\\
  \includegraphics[width=0.36\textwidth]{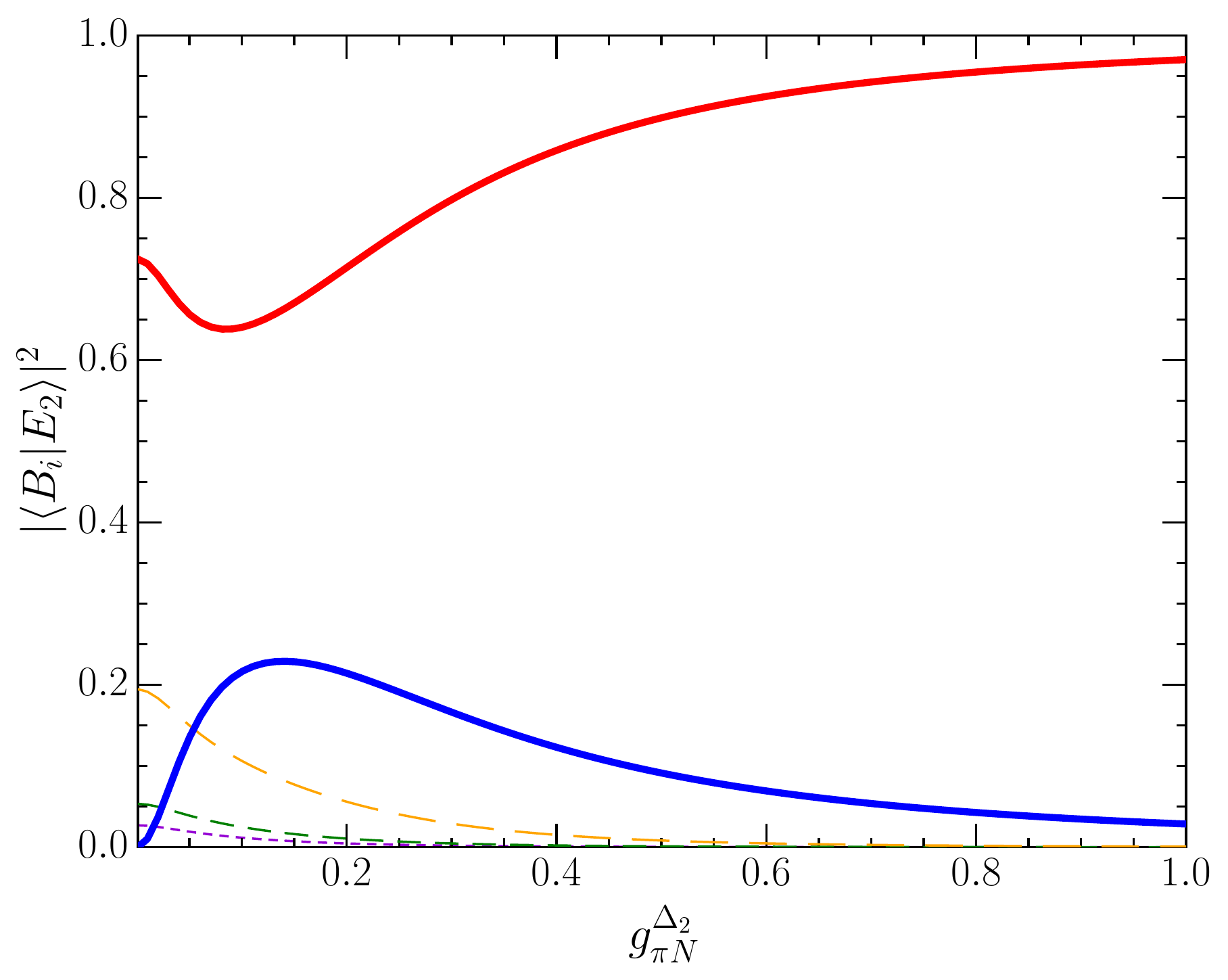}~\\
  \includegraphics[width=0.36\textwidth]{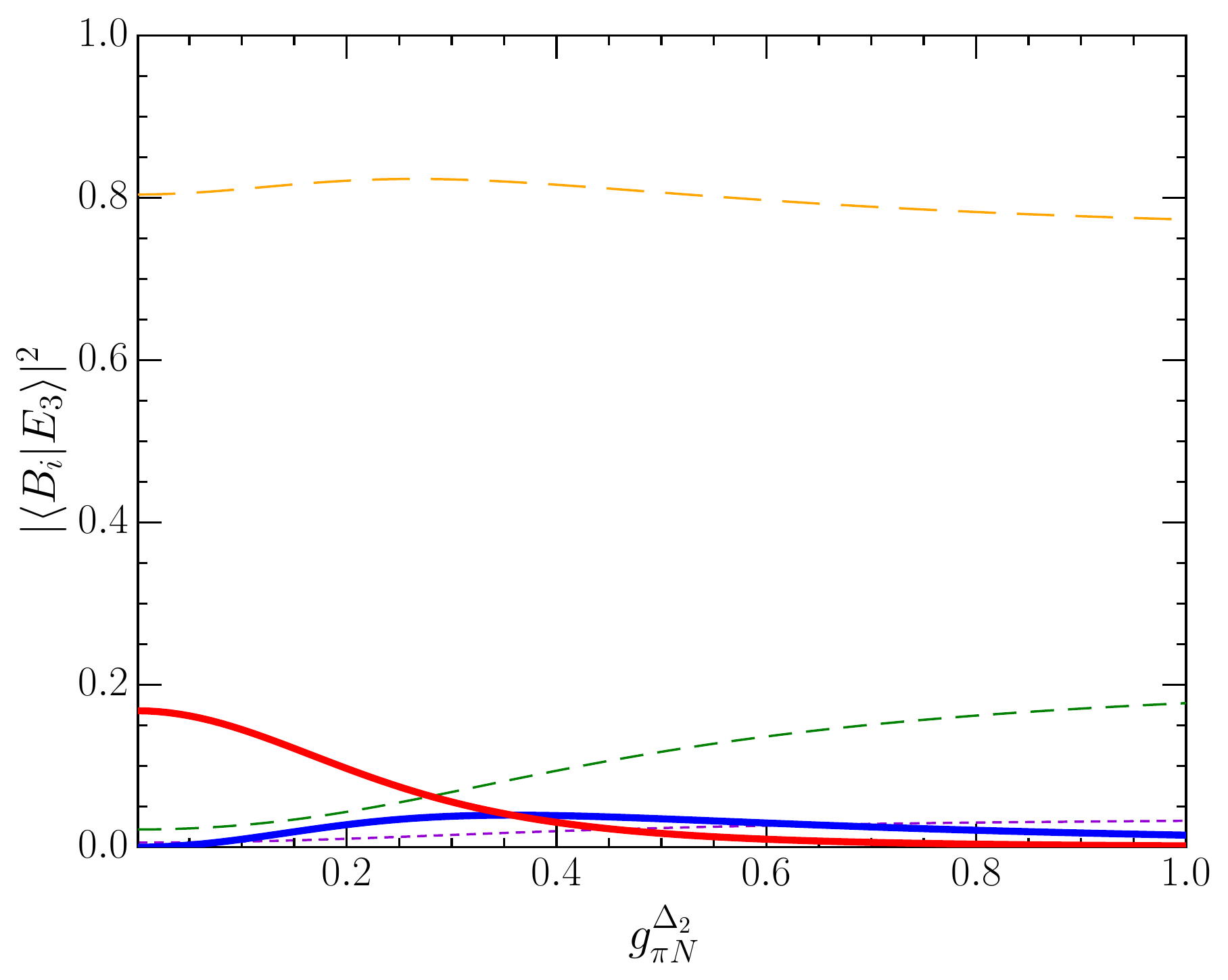}~\\
  \includegraphics[width=0.36\textwidth]{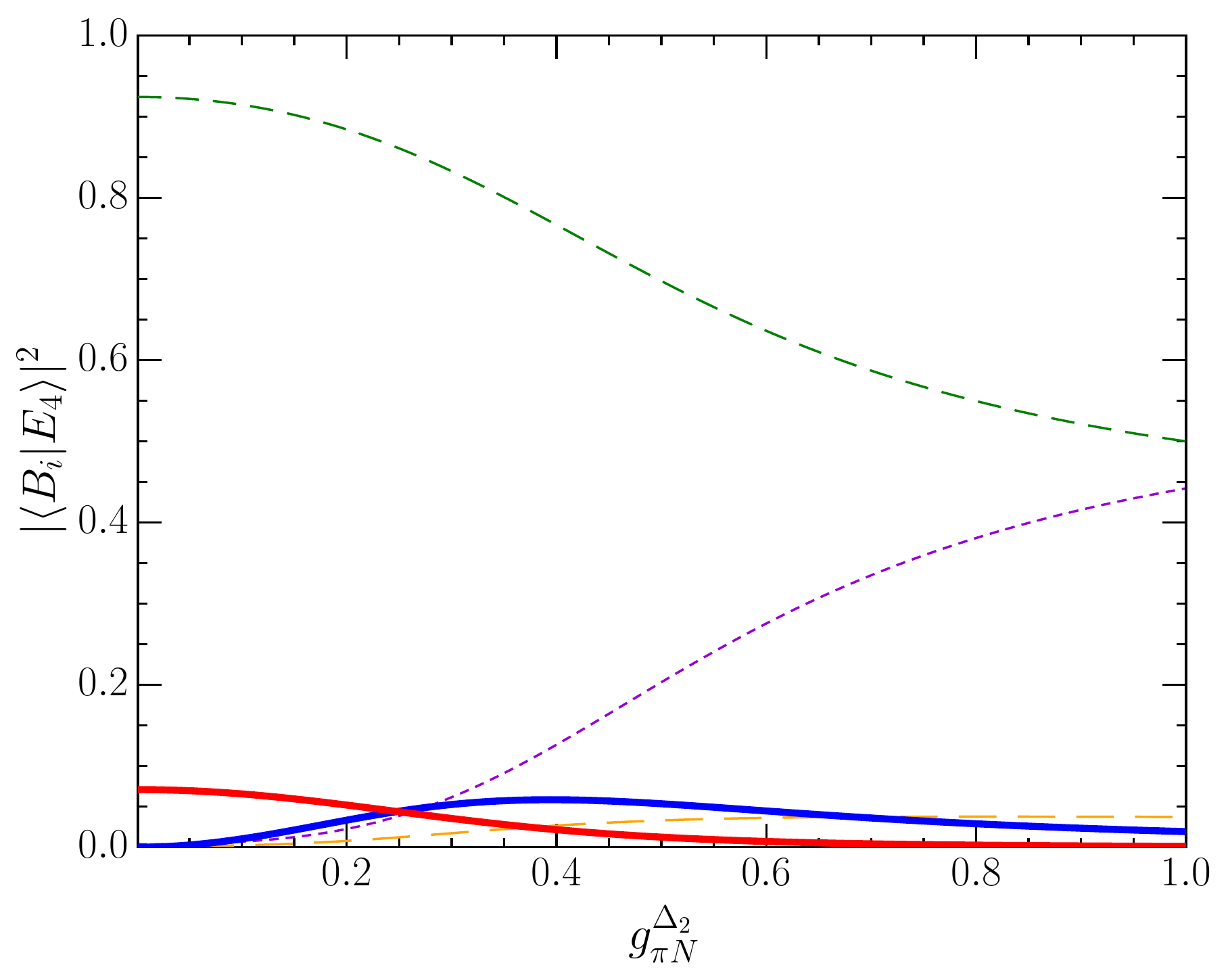}~
  \caption{Dependence of the finite-volume energy-eigenstate basis state structure on the coupling $g_{\pi N}^{\Delta_2}$, for $L = 3$ fm, where $m_{\Delta_2} = 1.2$ GeV.
    The label $\left| \bra{B_{i}}\ket{E_j} \right|^2$ denotes the contribution from each basis state $\ket{B_{i}}$ to the energy eigenvalues $\ket{E_{j}}$ from \fref{fig:2b1c_EvG_bare_L3fm_m1200}.
  Eigenvector components from higher momentum scattering states are summed in $\pi N(k\geq 3)$.}
  \label{fig:2b1c_eigenvectors_m1200}
\end{figure}

By constructing a finite-volume Hamiltonian that includes a second bare mass with $m_{\Delta_2} = 1.2$ GeV, lying below $m_{\Delta_1} = 1.359$ GeV, we are able to investigate whether the behaviour of the finite-volume eigenstates corresponds with the behaviour of the poles associated with each bare state as seen in \sref{subsec_inf_small}.
The resultant finite-volume energy spectrum as a function of the coupling of the 1.2 GeV bare state
to $\pi N$ scattering states is illustrated in \fref{fig:2b1c_EvG_bare_L3fm_m1200}.
In addition, the eigenstates with largest eigenvector components from each of the two bare states are highlighted in red and blue.
The eigenvector composition of the four lowest-lying eigenstates as a function of $g_{\pi N}^{\Delta_2}$ are given in \fref{fig:2b1c_eigenvectors_m1200}.
%
%

Considering the energy spectrum, we observe a similar behaviour to that corresponding to the pole movement discussed in \sref{subsec_inf_small}.
The state which initially describes the physical $\Delta$, represented by a pole at $1.210 - 0.050\,i$ GeV in an infinite volume, and by a solid red line in the finite-volume spectrum, moves from its initial position to the position of the first bare mass, at 1.359 GeV.
As this happens, the state becomes increasingly dominated by the contribution from the first bare state.
At a coupling of $g_{\pi N}^{\Delta_2} = 1$, 97\% of the eigenvector component for the first bare state resides in this eigenstate.

As this occurs, the ground state, which initially is the uncoupled light bare state, begins to mix with contributions from both $\pi N$ scattering states and the original bare state.
This rapidly reduces the value of the energy eigenvalue, corresponding with the movement of the pole in an infinite-volume, where it rapidly drops below the $\pi N$ threshold, becoming a bound state.
Considering the eigenvector composition of this state however, this reducing energy appears to be due to an increasing contribution in higher-momentum $\pi N$ scattering states.

Interestingly, the turning point observed in the pole movement, at $g_{\pi N}^{\Delta_1} = g_{\pi N}^{\Delta_2}$, is also observed in the eigenvectors for the two lowest-lying eigenstates for similarly sized values of $g_{\pi N}^{\Delta_2}$, further demonstrating the similar behaviour of the movement of poles in an infinite volume, with the eigenvector composition of eigenstates in a finite volume.

\section{Conclusion}
\label{sec:conclusion}
In this report, we have explored the effect of introducing a second quark-model like, bare basis state to a toy model extension of the $\Delta(1232)$ system.
As the $\Delta(1232)$ is the lowest-lying resonance, it is particularly suitable for such a study, where one can focus on the effects of multiple bare states, rather than the complex interplay of multiple scattering channels.

In particular, this study focused on the interplay between infinite and finite volume quantities, which vary in related manners as the coupling associated with the second bare state is increased.
In \sref{sec:inf}, we studied the behaviour of infinite-volume poles in the $T$-matrix, observing an interesting interplay between the pole positions of the two bare-state system.

With the introduction of a second, larger mass bare state, the initial pole associated with the
physical $\Delta(1232)$ quickly moves towards the real axis, taking on the role of a bound state.
The new pole associated with the second bare state has a similar behaviour, where it has the form of a resonance for small values of $g_{\pi N}^{\Delta_2}$, before tending towards the real axis at the mass of the original bare state.
The latter suggests an exchange in the roles of the bare states in generating the physical poles.
Upon introducing a second, lighter bare state, we observed the opposite behaviour.
Here, the new lower pole moves towards the position of a bound state below the $\pi N$ threshold, while the original pole moves to the position of the first bare mass, with a very small imaginary component.

In \sref{sec:fin}, we considered the behaviour of the finite-volume energy spectrum upon introducing a second bare state. We observed analogous behaviour to that found for the infinite-volume poles.
In both cases, the ground state rapidly moves below the $\pi N$ threshold and becomes dominated by eigenvector contributions from the second bare state.
By considering how the eigenvectors vary with $g_{\pi N}^{\Delta_2}$, we obtained new insight into the underlying physics driving the movements of the two poles in infinite-volume.
For $m_{\Delta_2} > m_{\Delta_1}$, we observed an exchange in the roles of the bare states, with $\Delta_{2}$ moving from $m_{\Delta_2}$ to the lowest lying state, and $\Delta_1$ moving to $m_{\Delta_1}$ at large $g_{\pi N}^{\Delta_2}$.

In both an infinite volume and a finite volume, we observed similar behaviour in the limit
where $g_{\pi N}^{\Delta_2}$ is approximately an order of magnitude larger
than $g_{\pi N}^{\Delta_1}$.
In this limit, both systems result in an unphysical bound state, and a state at the mass of the original bare state which appears to have relatively decoupled from $\pi N$ basis states.
It is clear that we require the bare states to have couplings of a similar order of magnitude to avoid the generation of unphysical poles and energy eigenstates.

While in the region of $g_{\pi N}^{\Delta_1} \sim g_{\pi N}^{\Delta_2}$, the eigenvector composition of the low-lying eigenstates indicates a significant contribution from each bare basis state.
In order words, the bare states mix to form the energy eigenstates.
The situation is reminiscent of the mixing of interpolating fields in isolating energy eigenstates in lattice QCD.

This strongly indicates the need for future studies with multiple bare basis states, as a system with two nearby poles indicates a significant mixing between the resonances associated with those poles when they are considered in a finite-volume.
One candidate for a study such as this involves the low-lying odd-parity nucleon resonances, the $N^{*}(1535)$ and $N^{*}(1650)$, where recent lattice QCD calculations \rref{Stokes:2019zdd} report magnetic moments in good agreement with the constituent quark model.
%

\section*{Acknowledgements}
This research was supported by the Australian Government Research Training Program Scholarship, and by supercomputing resources provided by the Phoenix HPC service at the University of Adelaide.
This research was undertaken with the assistance of resources from the National Computational
Infrastructure (NCI), provided through the National Computational Merit Allocation Scheme, and
supported by the Australian Government through Grant No.~LE190100021 and the University of Adelaide Partner Share.
This research was supported by the Australian Research Council through ARC Discovery Project Grants Nos. DP190102215 and DP210103706 (D.B.L.).
J.-J. Wu was support by the National Natural Science Foundation of China under Grants No.12175239, 12221005, and by the National Key R\&D Program of China under Contract No. 2020YFA0406400.

\appendix
\section{Solving the Scattering Equations}
As described in \sref{sec:theory}, the $T$-matrix may be solved from the coupled-scattering equations, given by
\begin{align}
  T_{\alpha\beta}&(k,k';E) = \tilde{V}_{\alpha\beta}(k,k';E) \nonumber\\
  &+ \sum_{\gamma}\, \int dq\, q^2\, \frac{\tilde{V}_{\alpha\gamma}(k,q;E)\, T_{\gamma\beta}(q,k';E)}{E - \omega_{\gamma}(q) + i\epsilon}\,,
\end{align}
where the sum over $\gamma$ sums over all scattering channels, with centre-of-mass energy $\omega_{\gamma}(q)$.
In addition, the coupled-channel potential $\tilde{V}_{\alpha\beta}(k,k';E)$ is given by
\begin{equation}
  \tilde{V}_{\alpha\beta}(k,k';E) = V_{\alpha\beta}(k,k') + \sum_{B_0} \frac{G_{\alpha}^{B_0\, \dagger}(k)\, G_{\beta}^{B_0}(k')}{E - m_{B_0}}\,,
\end{equation}
where the sum over $B_{0}$ considers all bare basis states.
In particular, in describing these scattering processes we consider separable potentials, which may be written as $V_{\alpha\beta}(k,k') = v_{\alpha\beta}\, f_{\alpha}(k)\, f_{\beta}(k')$, where $v_{\alpha\beta}$ is a dimensionless coupling strength.
In solving for the $T$-matrix, we separate the $T$-matrix into a sum of the background $T$-matrix, denoted $t_{\alpha\beta}(k,k';E)$, and the bare $T$-matrix, denoted $T_{\alpha\beta}^{\text{bare}}(k,k';E)$.

Considering at first the background $T$-matrix, this may be found as solutions of
\begin{align}
  t_{\alpha\beta}&(k,k';E) = V_{\alpha\beta}(k,k') \nonumber\\
  &+ \sum_{\gamma}\, \int dq\, q^2\, \frac{V_{\alpha\gamma}(k,q)\, T_{\gamma\beta}(q,k';E)}{E - \omega_{\gamma}(q) + i\epsilon}\,, \label{eq:tmat_background}
\end{align}
where we emphasise the use of the original two-particle potential $V_{\alpha\beta}(k,k')$, rather than the coupled-channel potential $\tilde{V}_{\alpha\beta}(k,k';E)$.
As we are considering interactions parametrised by separable potentials, this $t$-matrix is also separable \rref{Bagchi:1974zz}, and therefore may be written as $t_{\alpha\beta}(k,k';E) = f_{\alpha}(k)\, \tilde{t}_{\alpha\beta}(E)\, f_{\beta}(k')$.
Substituting this form into \eref{eq:tmat_background}, we may remove common factors of $f_{\alpha}(k)$ and $f_{\beta}(k')$, giving
\begin{equation}
  \tilde{t}_{\alpha\beta}(E) = v_{\alpha\beta} + \sum_{\gamma}\, \int dq\, q^2\, \frac{v_{\alpha\gamma}\, f_{\gamma}(q)^{2}\, \tilde{t}_{\gamma\beta}(E)}{E - \omega_{\gamma}(q) + i\epsilon}\,. \label{eq:ttilde_simple}
\end{equation}
This expression may be further simplified by defining the integral
\begin{equation}
  M_{\gamma}(E) = \int dq\, q^2\, \frac{f_{\gamma}(q)^2}{E - \omega_{\gamma}(q) + i\epsilon}\,, \label{eq:Mgamma}
\end{equation}
which may be solved using a Cauchy principle value integral.
In addition, by rewriting the components of \eref{eq:ttilde_simple} as matrices in channel-space, denoted by $\tilde{t}(E)$, $v$, and $M(E)$ respectively, we may rearrange \eref{eq:ttilde_simple} to give the matrix equation
\begin{equation} \label{eq:ttilde_matrix}
  \tilde{t}(E) = \left( 1 - v\, M(E) \right)^{-1}\, v\,.
\end{equation}
The full background $t$-matrix may therefore be found by $t_{\alpha\beta}(k,k';E) = f_{\alpha}(k)\, \tilde{t}_{\alpha\beta}(E)\, f_{\beta}(k')$.
In the case where one is interested in solving for the positions of dynamically-generated poles, these may be found by solving for complex energies $E$ such that $\det\left\{ \tilde{t}(E)^{-1} \right\} = 0$.

To construct the full $T$-matrix, we also require bare portion of the $T$-matrix, given by
\begin{align}
  T_{\alpha\beta}^{\text{bare}}&(k,k';E) \nonumber\\
  &= \sum_{B_0,B'_0}\, \mathcal{G}_{\alpha}^{B_0\, \dagger}(k;E) A_{B_0,B'_0}(E)\, \mathcal{G}_{\beta}^{B'_0}(k';E) \,,
\end{align}
Here we have defined two new expressions associated with the bare basis states, $\mathcal{G}_{\alpha}^{B_{0}}$ and $A_{B_{0},B_{0}'}(E)$.
For a system with only interactions between bare states and scattering states, and no background interactions, we simply have $\mathcal{G}_{\alpha}^{B_0}(k;E) = G_{\alpha}^{B_0}(k)$, independent of the on-shell energy $E$.
In the more general case, $\mathcal{G}_{\alpha}^{B_0}(k;E)$ is a modified potential describing how the bare states are dressed by the background interactions, denoted as the ``dressed potential'' in this text, and found as solutions of the integral equation
\begin{align}
  \mathcal{G}_{\alpha}^{B_0}&(k;E) = G_{\alpha}^{B_0}(k) \nonumber\\
  &+ \sum_{\gamma}\int dq\,q^2\, \frac{V_{\alpha\gamma}(k,q)\,\mathcal{G}_{\gamma}^{B_0}(q;E)}{E-\omega_\gamma(q)+i\epsilon}\,. \label{eq:mathcal_G_def}
\end{align}
However, this dressed potential is defined such that
\begin{equation}
  \frac{V_{\alpha\beta}(k,k')\,\mathcal{G}_{\beta}^{B_0}(k';E)}{E-\omega_\beta(k) + i\epsilon} = \frac{t_{\alpha\beta}(k,k';E)\,G_{\beta}^{B_0}(k')}{E-\omega_\beta(k) + i\epsilon}\,. \label{eq:mathcal_Gbeta}
\end{equation}
Utilising this property, as well as the separable nature of the background $t$-matrix, we may rewrite \eref{eq:mathcal_G_def} as
\begin{align}
  \mathcal{G}_{\alpha}^{B_0}&(k;E) = G_{\alpha}^{B_0}(k) \nonumber\\
                                  &+ \sum_{\gamma} \int dq\,q^2\, \frac{f_{\alpha}(k)\, \tilde{t}_{\alpha\gamma}(E)\, f_{\gamma}(q)\, G_{\gamma}^{B_0}(q)}{E-\omega_{\gamma}(q)+i\epsilon}\,. \label{eq:mathcal_G_sep}
\end{align}
For ease of calculation, we also define the integral
\begin{equation} \label{eq:g_f}
  g_{f,\gamma}^{B_0}(E) = \int dq\, q^2\, \frac{f_{\gamma}(q)\, G_{\gamma}^{B_0}(q)}{E - \omega_{\gamma}(q) + i\epsilon}\,,
\end{equation}
allowing \eref{eq:mathcal_G_sep} to be expressed as
\begin{equation}
  \mathcal{G}_{\alpha}^{B_0}(k;E) = G_{\alpha}^{B_{0}} + \sum_{\gamma} f_{\alpha}(k)\, \tilde{t}_{\alpha\gamma}(E)\, g_{f,\gamma}^{B_{0}}(E)\,.
\end{equation}

The second component of the bare $T$-matrix is the matrix of dressed bare states, given by
\begin{equation}
  A_{B_0,B'_0}(E) = \left[ \delta_{B_0,B'_0}\, (E - m_{B_0}) - \bar{\Sigma}_{B_0,B'_0}(E) \right]^{-1} \,,
\end{equation}
where $\bar{\Sigma}_{B_0,B'_0}(E)$ represents the sum of all one-loop self-energy interactions.
Within this component, we have self-energies associated with the interactions between bare basis states and two-particle states, of the form
\begin{equation}
  \Sigma_{B_{0},B_{0}'}(E) = \sum_{\gamma} \int dq\, q^2\, \frac{G_{\gamma}^{B_0}(q)\, G_{\gamma}^{B_0'}(q)}{E - \omega_\gamma(k) + i\epsilon}\,.
\end{equation}
In addition, we have self-energies describing the interactions between two-particle states, which we label background self-energies.
These self-energies are given by
\begin{equation}
  \Sigma_{B_0,B'_0}^{\text{I}}(E) = \sum_{\alpha,\beta} g_{f,\alpha}^{B_0}(E)\,\tilde t_{\alpha\beta}(E)\,g_{f,\beta}^{B'_0}(E)\,. \label{eq:selfEnergyI_full}
\end{equation}
In the case where one is interested in solving for the positions of any poles in the $T$-matrix associated with the presence of bare basis states, one need only to consider poles in the matrix of dressed states.
As a result, we may therefore find the positions of these poles for complex energies $E$ such that $\det\left\{ A_{B_{0},B_{0}'}(E)^{-1} \right\} = 0$.

Constructing the full $T$-matrix from $T_{\alpha\beta}(k,k';E) = t_{\alpha\beta}(k,k';E) + T_{\alpha\beta}^{\text{bare}}(k,k';E)$, one can also extract the scattering observables by considering the unitary $S$-matrix.
Considering an on-shell scattering process in channel $\alpha$, with centre-of-mass energy $E$, the on-shell momentum is given as the solution of $E = \sqrt{k_{\text{on}}^2 + m_{\alpha_{\text{M}}}^2} + \sqrt{k_{\text{on}}^2 + m_{\alpha_{\text{B}}}^2}$.
In this formalism, the $S$-matrix is parametrised such that the density of states in this channel is given by
\begin{equation}
  \rho_{\alpha}(E) = \frac{\omega_{\alpha_{\text{M}}}(k_{\text{on}})\, \omega_{\alpha_{\text{B}}}(k_{\text{on}})}{E}\, k_{\text{on}}^2\,,
\end{equation}
resulting in an $S$-matrix of the form
\begin{equation}
  S_{\alpha\beta}(E) = \delta_{\alpha\beta} - 2\pi i \sqrt{\rho_{\alpha}(E)\, \rho_{\beta}(E)}\, T_{\alpha\beta}(k_{\text{on}}, k_{\text{on}}; E)\,.
\end{equation}
As a result, at some on-shell energy $E$ the phase shift from channel $\alpha$ to channel $\beta$, and the inelasticity are found from $S(E) = \eta(E) \exp\left\{ 2i\, \delta_{\beta}(E) \right\}$, giving
\begin{align}
  \delta_{\beta}(E) &= \frac{1}{2} \atan\left\{ \frac{\mathrm{Im}\, S_{\alpha\beta}(E)}{\mathrm{Re}\, S_{\alpha\beta}(E)} \right\} \,, \nonumber\\
  \eta(E) &= \det \left\{ S(E) \right\}\,.
\end{align}
%
%

\section{Imaginary Component of the Poles} \label{app:imag_poles}
As seen in \fref{fig:2b1c_m16_poles} and \fref{fig:2b1c_m1200_poles}, when $g_{2}^{2} \gg g_{1}^{2}$ we observe a pole below the $\pi N$ threshold, and a pole at $E \approx m_{\Delta_{1}}$.
As the lower pole has moved below the $\pi N$ threshold, there are no available decay channels and therefore the pole must have no imaginary component.
For the larger mass, it is less clear why the imaginary component should become increasingly small as it approaches $m_{\Delta_1}$.
To probe this effect, we consider a simplified case as in Sec. III C of \refref{Abell:2021awi}, where all $\pi N \rightarrow \pi N$ interactions are turned off.
Additionally, for easy of notation we relabel the coupling strengths as $g_{i} = g_{\pi N}^{\Delta_{i}}$.
In this case, and substituting the form of $G_{\pi N}^{\Delta_{i}}(k)$ from \eref{eq:Gk}, the self-energies are given by
\begin{equation}
  \Sigma_{\Delta_{i}\Delta_{j}}(E) = g_{i} g_{j} \int dk\, \frac{k^2}{\omega_{\pi}(k)}\, \frac{u(k)^2}{E - \omega_{\pi N}(k) + i\epsilon}\,.
\end{equation}
As the integral in this expression is independent of the bare basis state, we label it $I_{\pi N}(E)$, giving $\Sigma_{\Delta_{i}\Delta_{j}}(E) = g_{i}\, g_{j}\, I_{\pi N}(E)$.

Using this notation, the matrix of bare states as defined in \eref{eq:Amat} is written as
\begin{align}
  &A(E)^{-1} = \nonumber\\
  &\begin{pmatrix}
    E - m_{\Delta_1} - g_{1}^{2}\, I_{\pi N}(E) & -g_{1}\, g_{2}\, I_{\pi N}(E) \\
    -g_{2}\, g_{1}\, I_{\pi N}(E) & E - m_{\Delta_2} - g_{2}^{2}\, I_{\pi N}(E)
  \end{pmatrix}\,.
\end{align}
Poles in the $T$-matrix are found for when the determinant of this matrix is zero, giving
\begin{align}
  &(E - m_{\Delta_1})(E - m_{\Delta_2}) - I_{\pi N}(E) \nonumber\\
  &\quad \times \left[ g_{1}^{2}(E - m_{\Delta_2}) + g_{2}^{2}(E - m_{\Delta_1}) \right] = 0\,.
\end{align}
By solving for energies $E$ which satisfy this equation, we obtain two poles, found as solutions of
\begin{align}
  2E &= m_{\Delta_1} + m_{\Delta_2} \nonumber\\
     &\qquad + g_{1}^{2} I_{\pi N}(E) + g_{2}^{2}I_{\pi N}(E) \pm \mu(E)\,, \label{eq:E_poles}
\end{align}
where we have defined the term
\begin{align}
  \mu(E)^{2} &= (m_{\Delta_{1}} - m_{\Delta_{2}} + g_{1}^{2}I_{\pi N}(E) - g_{2}^{2}I_{\pi N}(E))^2\nonumber\\
             &\qquad + 4\, g_{1}^{2}\, g_{2}^{2}\, I_{\pi N}(E)^2\,.
\end{align}
As we are interested in the case where $g_{2}^{2} \gg g_{1}^{2}$, we rearrange this expression into the form
\begin{align}
  \mu(E) &= g_{2}^{2}\, I_{\pi N}(E) \left\{ 1 + 2\, \frac{g_{1}^{2}}{g_{2}^{2}}\, \left( 1 - \frac{m_{\Delta_{1}} - m_{\Delta_{2}}}{g_{1}^{2}\, I_{\pi N}(E)} \right) \right.\nonumber\\
         &\qquad + \left. \frac{g_{1}^{4}}{g_{2}^{4}}\left( 1 + \frac{m_{\Delta_{1}} - m_{\Delta_{2}}}{g_{1}^{2}\, I_{\pi N}(E)} \right)^{2} \right\}^{1/2}\,.
\end{align}
Considering the expression inside the braces, we may expand about small $g_{1}^{2}/g_{2}^{2}$ up to $\mathcal{O}(g_{1}^{4}/g_{2}^{4})$, giving
\begin{align}
  \mu(E) &\approx g_{2}^{2}\, I_{\pi N}(E) \left\{ 1 + \frac{g_{1}^{2}}{g_{2}^{2}}\, \left( 1 - \frac{m_{\Delta_{1}} - m_{\Delta_{2}}}{g_{1}^{2}\, I_{\pi N}(E)} \right) \right. \nonumber\\
                &\qquad + \left. 2\, \frac{g_{1}^{4}}{g_{2}^{4}}\, \frac{m_{\Delta_{1}} - m_{\Delta_{2}}}{g_{1}^{2}\, I_{\pi N}(E)} \right\}\,, \nonumber\\
         &= -\left\{ m_{\Delta_{1}} - m_{\Delta_{2}} - I_{\pi N}(E) \left( g_{1}^{2} + g_{2}^{2} \right) \right. \nonumber\\
         & \qquad - \left. 2\, \frac{g_{1}^{2}}{g_{2}^{2}} \left( m_{\Delta_{1}} - m_{\Delta_{2}} \right) \right\}\,.
\end{align}

When substituting this expression into \eref{eq:E_poles}, taking the positive root of $\mu(E)$ produces the larger mass pole, and results in a mass of
\begin{equation}
  E = m_{\Delta_{1}} - \frac{g_{1}^{2}}{g_{2}^{2}}\, \left( m_{\Delta_{1}} - m_{\Delta_{2}} \right)\,.
\end{equation}
As the components of this expression are real, at $\mathcal{O}(g_{1}^{2}/g_{2}^{2})$ we therefore have no imaginary component for the larger mass pole.

Additionally, the pole mass tends towards $m_{\Delta_{1}}$ as $g_{1}^{2}/g_{2}^{2}$ tends to zero from the correct direction.
For $m_{\Delta_{1}} > m_{\Delta_{2}}$, as in \fref{fig:2b1c_m1200_poles}, the pole correctly tends to $m_{\Delta_{1}}$ from below, representing the movement from the original pole position at $E = 1.210 - 0.050\, i$ GeV.
For $m_{\Delta_{2}} > m_{\Delta_{1}}$, as in \fref{fig:2b1c_m16_poles}, the pole correctly tends to $m_{\Delta_{1}}$ from above, representing the movement from the new pole position beginning at $m_{\Delta_{2}}$.
%
%

\newpage
\bibliographystyle{elsarticle-num}
\bibliography{multiBareRefs}

\begin{thebibliography}{10}
\expandafter\ifx\csname url\endcsname\relax
  \def\url#1{\texttt{#1}}\fi
\expandafter\ifx\csname urlprefix\endcsname\relax\def\urlprefix{URL }\fi
\expandafter\ifx\csname href\endcsname\relax
  \def\href#1#2{#2} \def\path#1{#1}\fi

\bibitem{Luscher:1985dn}
M.~L{\"u}scher, Volume dependence of the energy spectrum in massive quantum
  field theories {{I}}. {{Stable}} particle states, Communications in
  Mathematical Physics 104~(2) (1986) 177--206.
\newblock \href {https://doi.org/10.1007/BF01211589}
  {\path{doi:10.1007/BF01211589}}.

\bibitem{Luscher:1986pf}
M.~L{\"u}scher, Volume dependence of the energy spectrum in massive quantum
  field theories {{II}}. {{Scattering}} states, Communications in Mathematical
  Physics 105~(2) (1986) 153--188.
\newblock \href {https://doi.org/10.1007/BF01211097}
  {\path{doi:10.1007/BF01211097}}.

\bibitem{Luscher:1990ux}
M.~L{\"u}scher, Two-particle states on a torus and their relation to the
  scattering matrix, Nuclear Physics B 354~(2) (1991) 531--578.
\newblock \href {https://doi.org/10.1016/0550-3213(91)90366-6}
  {\path{doi:10.1016/0550-3213(91)90366-6}}.

\bibitem{He:2005ey}
S.~He, X.~Feng, C.~Liu, Two particle states and the {{S}}-matrix elements in
  multi-channel scattering, JHEP 07 (2005) 011.
\newblock \href {http://arxiv.org/abs/hep-lat/0504019}
  {\path{arXiv:hep-lat/0504019}}, \href
  {https://doi.org/10.1088/1126-6708/2005/07/011}
  {\path{doi:10.1088/1126-6708/2005/07/011}}.

\bibitem{Lage:2009zv}
M.~Lage, U.-G. Meissner, A.~Rusetsky, A {{Method}} to measure the
  antikaon-nucleon scattering length in lattice {{QCD}}, Physics Letters B681
  (2009) 439--443.
\newblock \href {http://arxiv.org/abs/0905.0069} {\path{arXiv:0905.0069}},
  \href {https://doi.org/10.1016/j.physletb.2009.10.055}
  {\path{doi:10.1016/j.physletb.2009.10.055}}.

\bibitem{Bernard:2010fp}
V.~Bernard, M.~Lage, U.~G. Meissner, A.~Rusetsky, Scalar mesons in a finite
  volume, JHEP 01 (2011) 019.
\newblock \href {http://arxiv.org/abs/1010.6018} {\path{arXiv:1010.6018}},
  \href {https://doi.org/10.1007/JHEP01(2011)019}
  {\path{doi:10.1007/JHEP01(2011)019}}.

\bibitem{Guo:2012hv}
P.~Guo, J.~Dudek, R.~Edwards, A.~P. Szczepaniak, Coupled-channel scattering on
  a torus, Physical Review D88~(1) (2013) 014501.
\newblock \href {http://arxiv.org/abs/1211.0929} {\path{arXiv:1211.0929}},
  \href {https://doi.org/10.1103/PhysRevD.88.014501}
  {\path{doi:10.1103/PhysRevD.88.014501}}.

\bibitem{Hu:2016shf}
B.~Hu, R.~Molina, M.~D{\"o}ring, A.~Alexandru, Two-flavor {{Simulations}} of
  the \$\textbackslash{}rho(770)\$ and the {{Role}} of the
  \${{K}}\textbackslash{}bar {{K}}\$ {{Channel}}, Physical Review Letters
  117~(12) (2016) 122001.
\newblock \href {http://arxiv.org/abs/1605.04823} {\path{arXiv:1605.04823}},
  \href {https://doi.org/10.1103/PhysRevLett.117.122001}
  {\path{doi:10.1103/PhysRevLett.117.122001}}.

\bibitem{Li:2012bi}
N.~Li, C.~Liu, Generalized {{L}}\textbackslash{}"uscher {{Formula}} in
  {{Multi}}-channel {{Baryon}}-{{Meson Scattering}}, Physical Review D 87~(1)
  (2013) 014502.
\newblock \href {http://arxiv.org/abs/1209.2201} {\path{arXiv:1209.2201}},
  \href {https://doi.org/10.1103/PhysRevD.87.014502}
  {\path{doi:10.1103/PhysRevD.87.014502}}.

\bibitem{Hansen:2012bj}
M.~T. Hansen, S.~R. Sharpe, {Multiple-channel generalization of
  Lellouch-Lüscher formula}, PoS LATTICE2012 (2012) 127.
\newblock \href {http://arxiv.org/abs/1211.0511} {\path{arXiv:1211.0511}},
  \href {https://doi.org/10.22323/1.164.0127} {\path{doi:10.22323/1.164.0127}}.

\bibitem{Doring:2018xxx}
M.~D{\"o}ring, H.-W. Hammer, M.~Mai, J.-Y. Pang, A.~Rusetsky, J.~Wu, Three-body
  spectrum in a finite volume: The role of cubic symmetry, Physical Review D
  97~(11) (2018) 114508.
\newblock \href {https://doi.org/10.1103/PhysRevD.97.114508}
  {\path{doi:10.1103/PhysRevD.97.114508}}.

\bibitem{Hansen:2019nir}
M.~T. Hansen, S.~R. Sharpe, Lattice {{QCD}} and {{Three}}-particle {{Decays}}
  of {{Resonances}}, Annual Review of Nuclear and Particle Science 69~(1)
  (2019) 65--107.
\newblock \href {https://doi.org/10.1146/annurev-nucl-101918-023723}
  {\path{doi:10.1146/annurev-nucl-101918-023723}}.

\bibitem{Blanton:2019vdk}
T.~D. Blanton, F.~{Romero-L{\'o}pez}, S.~R. Sharpe, \${{I}} = 3\$ three-pion
  scattering amplitude from lattice {{QCD}}, Physical Review Letters 124~(3)
  (2020) 032001.
\newblock \href {http://arxiv.org/abs/1909.02973} {\path{arXiv:1909.02973}},
  \href {https://doi.org/10.1103/PhysRevLett.124.032001}
  {\path{doi:10.1103/PhysRevLett.124.032001}}.

\bibitem{Wu:2014vma}
J.-J. Wu, T.-S.~H. Lee, A.~W. Thomas, R.~D. Young, Finite-volume
  {{Hamiltonian}} method for coupled channel interactions in lattice {{QCD}},
  Physical Review C 90~(5) (2014) 055206.
\newblock \href {http://arxiv.org/abs/1402.4868} {\path{arXiv:1402.4868}},
  \href {https://doi.org/10.1103/PhysRevC.90.055206}
  {\path{doi:10.1103/PhysRevC.90.055206}}.

\bibitem{Thomas:1977ph}
A.~W. Thomas (Ed.), {Modern Three Hadron Physics}, Vol.~2, {Springer}, 1977.
\newblock \href {https://doi.org/10.1007/978-3-642-81070-1}
  {\path{doi:10.1007/978-3-642-81070-1}}.

\bibitem{Thomas:1982kv}
A.~W. Thomas, {Chiral Symmetry and the Bag Model: A New Starting Point for
  Nuclear Physics}, Adv. Nucl. Phys. 13 (1984) 1--137.
\newblock \href {https://doi.org/10.1007/978-1-4613-9892-9_1}
  {\path{doi:10.1007/978-1-4613-9892-9_1}}.

\bibitem{Hall:2013qba}
J.~M.~M. Hall, A.~C.~P. Hsu, D.~B. Leinweber, A.~W. Thomas, R.~D. Young,
  {Finite-volume matrix Hamiltonian model for a $\Delta \to N\pi$ system},
  Phys. Rev. D 87~(9) (2013) 094510.
\newblock \href {http://arxiv.org/abs/1303.4157} {\path{arXiv:1303.4157}},
  \href {https://doi.org/10.1103/PhysRevD.87.094510}
  {\path{doi:10.1103/PhysRevD.87.094510}}.

\bibitem{Abell:2021awi}
C.~D. Abell, D.~B. Leinweber, A.~W. Thomas, J.-J. Wu, {Regularization in
  nonperturbative extensions of effective field theory}, Phys. Rev. D 106~(3)
  (2022) 034506.
\newblock \href {http://arxiv.org/abs/2110.14113} {\path{arXiv:2110.14113}},
  \href {https://doi.org/10.1103/PhysRevD.106.034506}
  {\path{doi:10.1103/PhysRevD.106.034506}}.

\bibitem{Liu:2016uzk}
Z.-W. Liu, W.~Kamleh, D.~B. Leinweber, F.~M. Stokes, A.~W. Thomas, J.-J. Wu,
  {Hamiltonian effective field theory study of the $\mathbf{N^*(1440)}$
  resonance in lattice QCD}, Phys. Rev. D 95~(3) (2017) 034034.
\newblock \href {http://arxiv.org/abs/1607.04536} {\path{arXiv:1607.04536}},
  \href {https://doi.org/10.1103/PhysRevD.95.034034}
  {\path{doi:10.1103/PhysRevD.95.034034}}.

\bibitem{Wu:2017qve}
J.-j. Wu, D.~B. Leinweber, Z.-w. Liu, A.~W. Thomas, {Structure of the Roper
  Resonance from Lattice QCD Constraints}, Phys. Rev. D 97~(9) (2018) 094509.
\newblock \href {http://arxiv.org/abs/1703.10715} {\path{arXiv:1703.10715}},
  \href {https://doi.org/10.1103/PhysRevD.97.094509}
  {\path{doi:10.1103/PhysRevD.97.094509}}.

\bibitem{Hall:2014uca}
J.~M.~M. Hall, W.~Kamleh, D.~B. Leinweber, B.~J. Menadue, B.~J. Owen, A.~W.
  Thomas, R.~D. Young, {Lattice QCD Evidence that the
  \ensuremath{\Lambda}(1405) Resonance is an Antikaon-Nucleon Molecule}, Phys.
  Rev. Lett. 114~(13) (2015) 132002.
\newblock \href {http://arxiv.org/abs/1411.3402} {\path{arXiv:1411.3402}},
  \href {https://doi.org/10.1103/PhysRevLett.114.132002}
  {\path{doi:10.1103/PhysRevLett.114.132002}}.

\bibitem{Liu:2015ktc}
Z.-W. Liu, W.~Kamleh, D.~B. Leinweber, F.~M. Stokes, A.~W. Thomas, J.-J. Wu,
  Hamiltonian effective field theory study of the
  \$\textbackslash{}mathbf\{\vphantom\}{{N}}\^*(1535)\vphantom\{\}\$ resonance
  in lattice {{QCD}}, Physical Review Letters 116~(8) (2016) 082004.
\newblock \href {http://arxiv.org/abs/1512.00140} {\path{arXiv:1512.00140}},
  \href {https://doi.org/10.1103/PhysRevLett.116.082004}
  {\path{doi:10.1103/PhysRevLett.116.082004}}.

\bibitem{Thomas:2002sj}
A.~W. Thomas, {Chiral extrapolation of hadronic observables}, Nucl. Phys. B
  Proc. Suppl. 119 (2003) 50--58.
\newblock \href {http://arxiv.org/abs/hep-lat/0208023}
  {\path{arXiv:hep-lat/0208023}}, \href
  {https://doi.org/10.1016/S0920-5632(03)01492-0}
  {\path{doi:10.1016/S0920-5632(03)01492-0}}.

\bibitem{Young:2002ib}
R.~D. Young, D.~B. Leinweber, A.~W. Thomas, {Convergence of chiral effective
  field theory}, Prog. Part. Nucl. Phys. 50 (2003) 399--417.
\newblock \href {http://arxiv.org/abs/hep-lat/0212031}
  {\path{arXiv:hep-lat/0212031}}, \href
  {https://doi.org/10.1016/S0146-6410(03)00034-6}
  {\path{doi:10.1016/S0146-6410(03)00034-6}}.

\bibitem{Young:2002cj}
R.~D. Young, D.~B. Leinweber, A.~W. Thomas, S.~V. Wright, {Chiral analysis of
  quenched baryon masses}, Phys. Rev. D 66 (2002) 094507.
\newblock \href {http://arxiv.org/abs/hep-lat/0205017}
  {\path{arXiv:hep-lat/0205017}}, \href
  {https://doi.org/10.1103/PhysRevD.66.094507}
  {\path{doi:10.1103/PhysRevD.66.094507}}.

\bibitem{Li:2019qvh}
Y.~Li, J.-J. Wu, C.~D. Abell, D.~B. Leinweber, A.~W. Thomas, {Partial Wave
  Mixing in Hamiltonian Effective Field Theory}, Phys. Rev. D 101~(11) (2020)
  114501.
\newblock \href {http://arxiv.org/abs/1910.04973} {\path{arXiv:1910.04973}},
  \href {https://doi.org/10.1103/PhysRevD.101.114501}
  {\path{doi:10.1103/PhysRevD.101.114501}}.

\bibitem{Liu:2020foc}
Z.-W. Liu, J.-J. Wu, D.~B. Leinweber, A.~W. Thomas, {Kaonic Hydrogen and
  Deuterium in Hamiltonian Effective Field Theory}, Phys. Lett. B 808 (2020)
  135652.
\newblock \href {http://arxiv.org/abs/2003.09181} {\path{arXiv:2003.09181}},
  \href {https://doi.org/10.1016/j.physletb.2020.135652}
  {\path{doi:10.1016/j.physletb.2020.135652}}.

\bibitem{Li:2021mob}
Y.~Li, J.-j. Wu, D.~B. Leinweber, A.~W. Thomas, {Hamiltonian effective field
  theory in elongated or moving finite volume}, Phys. Rev. D 103~(9) (2021)
  094518.
\newblock \href {http://arxiv.org/abs/2103.12260} {\path{arXiv:2103.12260}},
  \href {https://doi.org/10.1103/PhysRevD.103.094518}
  {\path{doi:10.1103/PhysRevD.103.094518}}.

\bibitem{Guo:2022hud}
D.~Guo, Z.-W. Liu, {Pion photoproduction off nucleon with Hamiltonian effective
  field theory}, Phys. Rev. D 105~(11) (2022) 114039.
\newblock \href {http://arxiv.org/abs/2201.11555} {\path{arXiv:2201.11555}},
  \href {https://doi.org/10.1103/PhysRevD.105.114039}
  {\path{doi:10.1103/PhysRevD.105.114039}}.

\bibitem{site:SAID}
{INS Data Analysis Center}, \url{http://gwdac.phys.gwu.edu/}, online, Solution
  W108.

\bibitem{Workman:2012hx}
R.~L. Workman, R.~A. Arndt, W.~J. Briscoe, M.~W. Paris, I.~I. Strakovsky,
  {Parameterization dependence of T matrix poles and eigenphases from a fit to
  $\pi$N elastic scattering data}, Phys. Rev. C 86 (2012) 035202.
\newblock \href {http://arxiv.org/abs/1204.2277} {\path{arXiv:1204.2277}},
  \href {https://doi.org/10.1103/PhysRevC.86.035202}
  {\path{doi:10.1103/PhysRevC.86.035202}}.

\bibitem{ParticleDataGroup:2022pth}
R.~L. Workman, et~al., {Review of Particle Physics}, PTEP 2022 (2022) 083C01.
\newblock \href {https://doi.org/10.1093/ptep/ptac097}
  {\path{doi:10.1093/ptep/ptac097}}.

\bibitem{Stokes:2019zdd}
F.~M. Stokes, W.~Kamleh, D.~B. Leinweber, {Elastic Form Factors of Nucleon
  Excitations in Lattice QCD}, Phys. Rev. D 102~(1) (2020) 014507.
\newblock \href {http://arxiv.org/abs/1907.00177} {\path{arXiv:1907.00177}},
  \href {https://doi.org/10.1103/PhysRevD.102.014507}
  {\path{doi:10.1103/PhysRevD.102.014507}}.

\bibitem{Bagchi:1974zz}
B.~Bagchi, B.~Mulligan, {Off-shell T matrices for a class of separable nonlocal
  potentials}, Phys. Rev. C 10 (1974) 2197--2205.
\newblock \href {https://doi.org/10.1103/PhysRevC.10.2197}
  {\path{doi:10.1103/PhysRevC.10.2197}}.

\end{thebibliography}
\end{document}